\documentclass[reprint,aps,rmp,amsmath,amssymb,longbibliography,superscriptaddress,floatfix]{revtex4-1}

\usepackage{xcolor}
\usepackage{graphicx}
\usepackage{amsmath}
\usepackage{amssymb}
\usepackage{amsfonts}
\usepackage{units}
\usepackage{bm}
\usepackage{hyperref}
\usepackage{enumerate}
\usepackage{longtable}
\usepackage{totcount}

\newtotcounter{citnum}
\def\oldbibitem{} \let\oldbibitem=\bibitem
\def\bibitem{\stepcounter{citnum}\oldbibitem}

\DeclareMathOperator{\Tr}{Tr}

\renewcommand{\vec}[1]{\bm{#1}}
\newcommand{\talpha}{\alpha}

\begin{document}

\title{Colloquium: Nonequilibrium effects in superconductors with a spin-splitting field}

\author{F. Sebastian Bergeret}
\email{sebastian\_bergeret@ehu.eus}
\affiliation{Centro de Fisica de Materiales (CFM-MPC), Centro Mixto CSIC-UPV/EHU, Manuel de Lardizabal 4, E-20018 San Sebastian, Spain}
\affiliation{Donostia International Physics Center (DIPC), Manuel de Lardizabal 5, E-20018 San Sebastian, Spain}

\author{Mikhail Silaev}

\affiliation{University of Jyvaskyla, Department of Physics and Nanoscience Center,
P.O. Box 35 (YFL), FI-40014 University of Jyv\"askyl\"a, Finland }

\author{Pauli Virtanen}

\affiliation{NEST, Istituto Nanoscienze-CNR and Scuola Normale
  Superiore, I-56127 Pisa, Italy }

\author{Tero T. Heikkil\"a}
\email{Tero.T.Heikkila@jyu.fi}

\affiliation{University of Jyvaskyla, Department of Physics and Nanoscience Center,
P.O. Box 35 (YFL), FI-40014 University of Jyv\"askyl\"a, Finland }

\begin{abstract}
We review the recent progress in understanding  the { properties of}   spin-split superconductors  under non-equilibrium conditions.  Recent experiments and theories demonstrate a rich variety of  transport phenomena occurring in  devices based on such materials  that    suggest direct   applications in thermoelectricity,  low-dissipative spintronics,  radiation detection and sensing.    We discuss different experimental situations and present {a theoretical} framework based on quantum kinetic equations. Within this framework we provide  an accurate  description of  the non-equilibrium distribution of charge, spin and energy, which are the relevant non-equilibrium modes, in different hybrid structures.  
We also review  experiments on  spin-split superconductors and show how  transport measurements reveal the properties of the non-equilibrium  modes and their  mutual coupling. We discuss in detail {spin injection and diffusion and very large thermoelectric effects in spin-split superconductors.}

\end{abstract}
\pacs{}
\maketitle

\tableofcontents

\section{Introduction\label{sec-introduction}}

Ferromagnetism and spin-singlet superconductivity are antagonist orders and hardly coexist in bulk materials. However, hybrid nanostructures allow for the possibility of combining the two phenomena via mutual proximity effects. The combination leads to the {emergence}  of novel features not present in either system alone.  We can make a distinction  among those characteristics affecting the spectral properties of the materials, showing up when the probed systems are in equilibrium, and those related to nonequilibrium phenomena. The emphasis of our text is in the latter phenomena, especially related to steady-state currents or voltages applied across the structures.

Both superconductors and ferromagnets are examples of electron systems with spontaneously broken symmetries, and thereby characterized by order parameters. The order parameter for a conventional spin singlet superconductor is the amplitude of (Cooper) pairing between electrons in states with opposite spins and momenta \cite{bardeen1957theory}. The presence of this complex pairing amplitude $F$ leads to two characteristic features of conventional superconductivity \cite{TinkhamBook,deGennes:566105}: An equilibrium supercurrent that is proportional to the gradient of the phase of $F$ and that can be excited without voltage, and to the quasiparticle spectrum exhibiting an energy gap proportional to the absolute value of $F$. The resulting density of states (DOS, Eq.~\eqref{eq:split_Dos} for $h_{\rm eff}=0$) is strongly energy dependent and results into a non-linear nonequilibrium response of superconductors.

The main defining feature of ferromagnets is the broken spin-rotation symmetry into the direction of magnetization, and the associated exchange energy $h$ that splits the spin up and down spectra. This also leads to a strong spin dependence (spin polarization) of the observables related to ferromagnets.

There are 
 two mechanisms that  prevent most of the ferromagnetic materials from becoming superconducting. One of them is  the orbital effect due to the intrinsic magnetic field in ferromagnets. 
 When this field exceeds a certain  critical value, superconductivity is suppressed
 \cite{Ginzburg:1957}.
 The second mechanism is the paramagnetic effect \textcite{Chandrasekhar:1962,PhysRevLett.9.266,saint1969type}.  
 This is due to the intrinsic exchange field of the ferromagnet that shows up as  a splitting of the energy levels of spin-up and spin-down electrons 
and hence prevents the formation of  Cooper pairs. We focus here on the regime where this spin-splitting field is present, but not yet too large to kill superconductivity.

In superconductors the spin-splitting field  can be generated either due to the
Zeeman effect in magnetic field or as a  result of the exchange interaction between the electrons forming Cooper pairs and  those which determine the magnetic order.  Such fields can lead to drastic modifications of the ground state of  a spin-singlet superconductor.  The best-known example is the formation of the spatially inhomogeneous superconducting state predicted by 
 \textcite{fulde_ferrell.1964} and \textcite{larkin_ovchinnikov.1964} and dubbed as FFLO. Although extensively studied in the literature, the FFLO phase 
 only takes place in a narrow parameter window and therefore its experimental realization is challenging.

\begin{figure*}
\includegraphics[width=1.8\columnwidth]{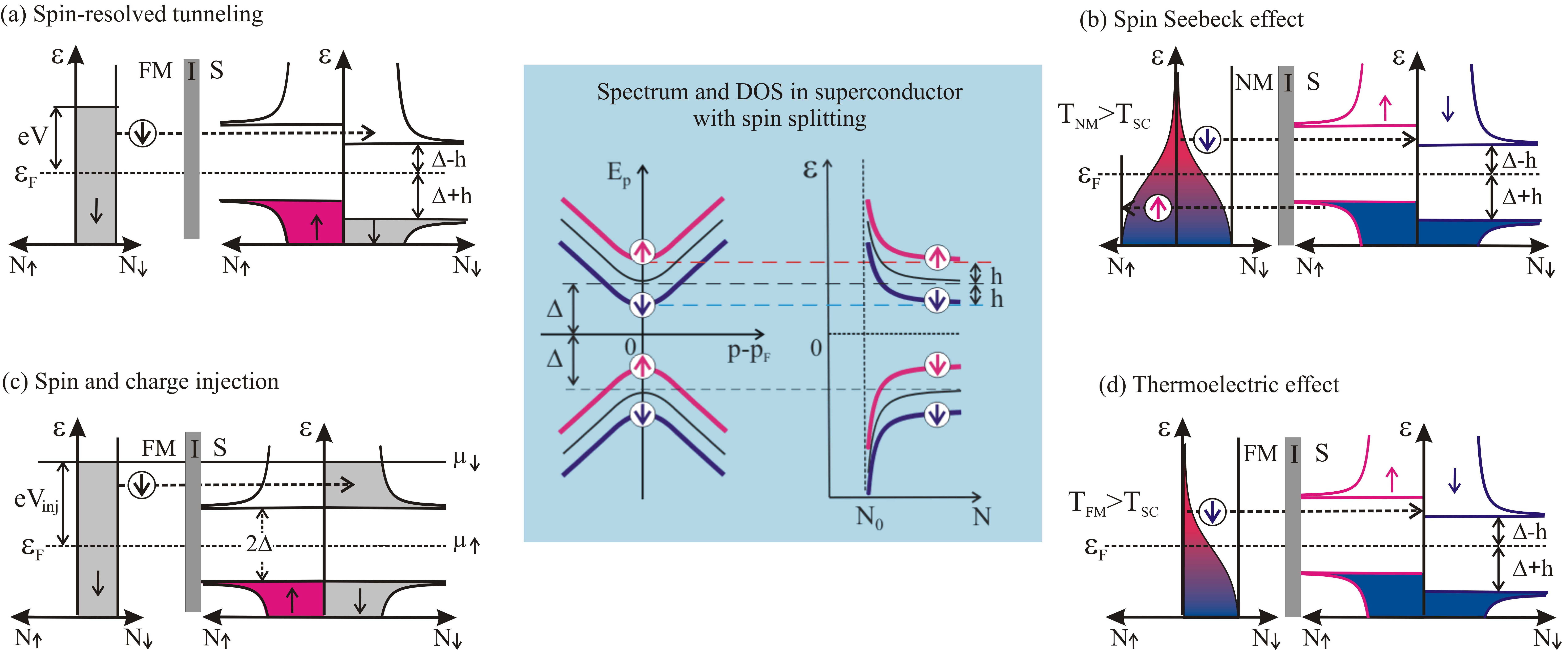}
\caption{
Central panel: quasiparticle spectrum and density of states in a superconductor with spin splitting, {\bm  $N_0$ is the normal metal DOS.}
(a-d) Schematic pictures of  various nonequilibrium phenomena  occurring at  normal metal/insulator/superconductor (NM/I/S) and ferromagnetic/insulator/superconductor (FM/I/S) interfaces discussed in this review. For clarity we show the limit of half-metallic FM with  $N_\uparrow = 0$.  
(a) Spin-resolved tunneling from a ferromagnetic metal  to a spin-split superconductor that leads to the spin valve effect, {\it i.e.}, the charge current in the parallel  magnetic configuration is different from that in the anti-parallel one. 
(c) Creation of spin and charge accumulation in the voltage biased FM/S junction. 
(b,d) Schematic picture of thermally excited currents in NM/S and FM/S junctions with a spin-split superconductor.
(b) Spin Seebeck effect in NM/S junction: A pure spin current is generated by the temperature bias between a spin-split superconductor at temperature $T_{SC}$ and a normal metal at temperature $T_{NM}>T_{SC}$. 
(d) Thermoelectric effect in a FM/I/S junction: 
Here the spin current is partially converted to the charge current due to the spin-dependent density of states in the ferromagnet.  
\label{fig:SemiconductorPicture}
}
\end{figure*}

Other more robust phenomena related to the spin-splitting fields in superconductors have their origin in the quasiparticle spectrum modification. In the central panel of Fig.~\ref{fig:SemiconductorPicture} we show  the resulting spin-split density of states. 
This was first explored experimentally by  \textcite{PhysRevLett.25.1270,meservey1975tunneling} through the spin-valve effect in the superconductor/ferromagnet (Al/Ni) tunnel junctions (Fig. \ref{fig:SemiconductorPicture}a).  
  In these experiments the magnetic field was applied  in the plane of a thin superconducting film,  such that the paramagnetic effect dominates. The spin-split DOS was utilized to determine the spin polarization of  an adjacent ferromagnet
  \cite{ PhysRevLett.26.192, tedrow1973spin,PhysRevB.16.4907, PhysRevB.22.1331,meservey1994spin}. The basis  of  this spin-valve effect
is the spin-resolved tunneling into the superconductor with spin splitting, shown  in  Fig.~\ref{fig:SemiconductorPicture}a. This schematic picture illustrates how by properly tuning the voltage across the junction,  the  electronic transport is dominated by only one of the spin species.  That results in peculiar  asymmetric differential conductance curves $dI/dV (V)\neq  dI/dV (-V)$  observed in experiments and revealing the spin polarization. This idea has  been  used more recently to probe the  spatially resolved spin polarization  of different magnetic materials by means of scanning tunneling microscopy  with spin-split superconducting tips \cite{Eltschka2014,Eltschka2015}.  Similar effects  can also arise in thin superconducting films by the magnetic proximity effect from an adjacent ferromagnetic material \cite{PhysRevLett.56.1746}. In such a case, the spin splitting of the density of states can be observed for  small magnetic fields or even at zero field, as discussed  in Sec.~\ref{sec-superwithh}.

The combination of spin-splitting fields with strong spin-orbit interaction in superconducting nanowires  has also raised considerable interest as a platform for realizing topological phases and Majorana fermions,   with possible applications in topological quantum computation \cite{aasen2016milestones}.  Although these effects are beyond the focus of this review, the physics discussed below may help in understanding transport properties of the devices studied in that context.

Due to the different nature of their broken symmetry, combining superconductors (S) and ferromagnets (FM) in hybrid structures leads to a multitude of effects where magnetism affects superconductivity and vice versa. Some of these effects show up already in equilibrium properties, especially studied in the context of  proximity effects in superconducting/metallic ferromagnet hybrids and reviewed for example by \textcite{RevModPhys.77.935} and \textcite{Bergeret2005}. The latter usually focus on the unusual behavior of  Cooper pairs leaking from a superconductor into a  metallic ferromagnet generating, for example, oscillating pair wave functions analogous to the FFLO state \cite{Buzdin:1982,demler1997superconducting} and long-range spin triplet correlations  \cite{bergeret2001long} induced by the coupling between  the  intrinsic exchange field of  the  ferromagnet and  the leaked superconducting condensate \cite{bergeret2001long}. These effects manifest themselves in measurable equilibrium effects, such  as
   the density of states and critical temperature oscillations in S/FM bilayers \cite{jiang1996superconducting,kontos2001inhomogeneous},
      triplet spin valve effects in  the critical temperature of FM/S/FM structures \cite{PhysRevX.5.021019}, 
      and  unusual Josephson effects in SC/FM/SC junctions \cite{ryazanov2001coupling,Singh2016}.
Inversely,  a  magnetic proximity effect can arise when the triplet pairs, created in the FM region, leak back into the superconductor in a FM/S metallic bilayer,  generating a non-vanishing magnetic moment in the SC within a coherence length $\xi_s$ from the SM/FM interface  \cite{bergeret2004induced}.
  
In contrast to these equilibrium proximity effects,  here 
  we focus  on  nonequilibrium properties of a superconducting material 
with a  built-in spin-splitting field. The interest in studying such systems has been  intensified recently due to the 
 technological advances which allow for a controllable generation of spin splitting 
  in thin superconducting films either by applying  an  external in-plane magnetic 
  field \cite{Hubler2012a,Quay2013} or by an adjacent ferromagnetic insulator. Structures with insulating FMs avoid the proximity effect 
  suppressing superconductivity. 
   Such nonequilibrium properties are studied by applying currents or voltages across the structures. The focus of our Colloquium is on steady-state nonequilibrium effects with time independent driving fields, but we also mention works studying alternating current (ac) responses.

Often the nonequilibrium effects  can survive to much higher distances than $\xi_s$, as their decay scales are determined via the various inelastic and spin-flip scattering lengths. Moreover, they can be studied at a weak tunneling contact to ferromagnets, making the analysis in some cases more straightforward than in proximity experiments. Nonequilibrium properties are related to the deviation of the electron distribution function from its equilibrium form, which leads to a nonequilibrium distribution (imbalance) of charge, energy or spin degrees of freedom. We refer to these different types of deviations from equilibrium as nonequilibrium modes.\footnote{The term "mode" here refers to the changes of the electron distribution function with respect to its equilibrium form. It should be distinguished from collective modes such as the \textcite{carlson1973superconducting} or the amplitude mode \cite{higgs64} that affect the response of superconductors at temperatures close to the critical temperature or at high frequencies.} Specifically, 
we explore the coupling  between  these modes  in superconductors with a spin-splitting field, and discuss unusually strong thermoelectric response and long-range spin signals.

The above mentioned ability to characterize the spin polarized Fermi surface of metallic magnets with the help of spin-split superconductors  has a direct connection with spintronics, and in particular with the search for  spin valves with larger efficiencies than in the  structures exhibiting large magnetoresistance \cite{PhysRevLett.61.2472,PhysRevB.39.4828,moodera1995large}. Indeed,  a superconductor with a spin-splitting field has also an intrinsic energy dependent spin polarization around the Fermi level. This allows for studying different  spintronic effects in a setting of a controllable non-linearity arising from the superconducting gap. Some of these effects are schematically shown  in Fig.~\ref{fig:SemiconductorPicture}. This review explains those phenomena in detail.

 In normal metals and superconductors a  spin accumulation, or spin imbalance, can be  created by injection of a charge current from a ferromagnetic electrode \cite{PhysRevLett.55.1790,van1987boundary,jedema:345,Johnson1994,gu2002direct,PhysRevB.71.144513,Takahashi2003,poli2008spin}. 
This state is characterized by the excess population in one of the spin subbands, determined by the balance between spin injection and relaxation or spin diffusion rates. In normal metals 
the nonequilibrium spin imbalance decays due to spin-flip scattering at typical distances  of several hundreds of nanometers.  
In the superconducting state, at low temperatures $k_B T\ll \Delta$ the injection of any amount of carriers just above the energy gap shifts the  chemical potential of quasiparticles rather strongly due to the large amount of quasiparticles at the gap edge  [Fig.~\ref{fig:SemiconductorPicture}(c)]. 
 This leads to a strong  spin signal in SF junctions \cite{Takahashi2003,poli2008spin}.   

The spin relaxation length  in normal metals 
depends only   weakly  on the temperature $T$. 
In  the superconducting state, however, the scattering length is drastically modified with $T$.  According to the first theory and experiments on spin injection in superconductors, the spin relaxation length  was found to be reduced compared to the normal state \cite{morten2004spin,poli2008spin}. 
However, subsequent experiments  showed, contrary to expectations, an increase of the spin decay length  \cite{Hubler2012a,Quay2013}.
It is now understood that these findings  can only be explained  by taking into account the  spin-splitting field inside the superconductor \cite{silaev2015long,krishtop2015-nst,bobkova2016injection,bobkova2015}. 
Due to this field, as shown in Sec.~\ref{sec:noneq_quasiclas}, 
it is necessary to take into account four types of nonequilibrium modes describing spin, charge, energy, and spin-energy imbalances. These modes provide the natural generalization of the charge and energy imbalances introduced  by \textcite{schmid1975}.  In Sec.~\ref{spininjection} we show how the spin-splitting field couples pairwise these modes: charge to spin energy and spin to energy. Such a coupling leads to striking effects.  For example,  the coupling between the spin and energy modes leads to the long-range spin-accumulation observed in the experiments by \textcite{Hubler2012a,Quay2013}.  As we show in Sec.~\ref{spininjection} this long-range effect  is related to the fact that the energy mode can only relax via inelastic processes which at low temperatures are rare.

The coupling between different modes shows up also in tunnel contacts with  spin-split superconductors. Because the spin-splitting field shifts the spin-resolved DOS away from the chemical potential of the superconductor, the system exhibits a strong spin-dependent electron-hole asymmetry. The spin-averaged density of states is still electron-hole symmetric, and therefore does not violate fundamental symmetries of the (quasiclassical) superconducting state. This spin-resolved electron-hole asymmetry leads to a large {\it spin Seebeck effect} shown schematically in Fig.~\ref{fig:SemiconductorPicture}b and discussed in Sec.~\ref{subs:spinseebeck}. A temperature difference across a tunneling interface between a normal metal and a spin-split superconductor drives a pure spin current between the electrodes, without transport of charge. If one of the electrodes is small so that the spin injection rate is large or comparable to the rate for spin relaxation, a spin accumulation forms in this electrode. 

However, it was noticed in several recent  works \cite{Ozaeta2014a,Machon2014,Machon2013,kalenkov12} that in certain situations the relevant observables are not spin-averaged, resulting in  an effective electron-hole asymmetry showing up also in the charge current. The spin components are weighted differently in a setup consisting of the spin-filter junction connected to the spin-split superconductor \cite{Ozaeta2014a}, shown schematically in Fig.~\ref{fig:SemiconductorPicture}d. As a result of this effective electron-hole symmetry breaking, the system exhibits  a very large  thermoelectric effect. This is discussed  in Sec.~\ref{thermoel}.

The main body of the review   is organized as follows. In Sec.~\ref{sec-superwithh} we describe spin-split superconductors and give an overview of the quasiclassical theory that can be used for describing both their equilibrium and nonequilibrium properties. In Sec.~\ref{sec:modes} we describe the nonequilibrium modes in superconducting systems driven out of equilibrium in terms of the quasiclassical formalism. Section \ref{spininjection} focuses  on the spin injection and diffusion in superconducting systems, and reviews experiments performed to detect spin and charge imbalance in superconductors with and without spin-splitting. In Sec.~\ref{thermoel} we describe the giant thermoelectric response of a system exhibiting spin-polarized tunneling into a superconductor with a spin-splitting field.  Finally, we present our conclusions and an outlook on possible future developments  in the field in Sec.~\ref{sec:outlook}. A longer version of this review, along with comprehensive technical detail, can be found at \cite{bergeret2017nonequilibrium}.

\section{Superconductor with an exchange field \label{sec-superwithh}}
\label{sec:SCwithh}

The main focus of this colloquium is on superconductors with a spin-split density of states (DOS). As discussed in the introduction such a splitting can originate either  by  an external magnetic field \cite{PhysRevLett.25.1270} or by the exchange field induced by an adjacent ferromagnetic insulator \cite{PhysRevLett.56.1746}. The split DOS  was observed  in  spectroscopy experiments \cite{PhysRevLett.25.1270, PhysRevLett.26.192,PhysRevB.16.4907,PhysRevB.22.1331,Hao:1990,PhysRevLett.106.247001}.

Formally, the normalized  DOS of a spin-split Bardeen-Cooper-Schrieffer (BCS)
 superconductor is expressed  
 as the sum of the DOS of each spin species, $N=N_\uparrow+N_\downarrow$,
\begin{equation}
N=\frac{1}{2}{\rm Re}\frac{ \varepsilon+h_{\rm eff}}{\sqrt{\left(\varepsilon+h_{\rm eff}\right)^{2}-\Delta^{2}}}+\frac{1}{2}{\rm Re}\frac{\varepsilon-h_{\rm eff}}{\sqrt{\left(\varepsilon-h_{\rm eff}\right)^{2}-\Delta^{2}}}\;, \label{eq:split_Dos}
\end{equation}
where $\pm h_{\rm eff}$ is the effective spin-splitting
field. Equation (\ref{eq:split_Dos}) is a simplified description
because it  does not take into account  the effect of  magnetic impurities  or   spin-orbit coupling  (SOC) \cite{meservey1994spin} discussed below. Often inelastic processes are described by $\varepsilon \mapsto \varepsilon + i \Gamma$, where $\Gamma$ is the \textcite{dynes84}
parameter.

In the case when the exchange field is induced by an adjacent ferromagnetic insulator (FI) there is  no need  of applying an  external  magnetic field \cite{PhysRevLett.56.1746,Hao:1990,Wolf2014,PhysRevLett.106.247001,Moodera_review,spin_filter_Blamire}. Microscopically, the spin splitting originates from the exchange interaction  between conduction electrons and the magnetic moments of the FI localized at the S/FI interface \cite{PhysRevB.38.8823,khusainov1996indirect,Kushainov_review}. 
 The ferromagnetic ordering in the FI is due to a direct 
 exchange coupling between  the localized  magnetic moments.  
In usual FIs the direct coupling is strong enough that one can assume 
that the magnetic configuration of the FI is {only} weakly affected by the superconducting state \cite{buzdin1988ferromagnetic,Bergeret2000}.

{The modification of the DOS is non-local} and survives over distances away from the FI/S interface of the order of the   coherence length  $\xi_{s}$ \cite{PhysRevB.38.8823,bergeret2004induced}.  {If  the thickness $d$ of the S film is  much smaller
than  $\xi_{s}$, the spin splitting  can be assumed as  homogeneous across the film. Thus the density of states can be approximated by  Eq.~(\ref{eq:split_Dos}) with an effective  exchange field  ${\bf h}_{\rm eff}\approx J_{\rm ex}\langle{\bf S_r}\rangle a/d$ \cite{deGennes1966coupling,khusainov1996indirect,PhysRevB.38.8823}, where $a$ is the characteristic distance between the localized spins, $J_{\rm ex}$ is the exchange coupling between conduction electrons and localized moments, and $\langle{\bf S_r}\rangle$ is the average of the latter}. 

\begin{figure}
\includegraphics[width=\columnwidth]{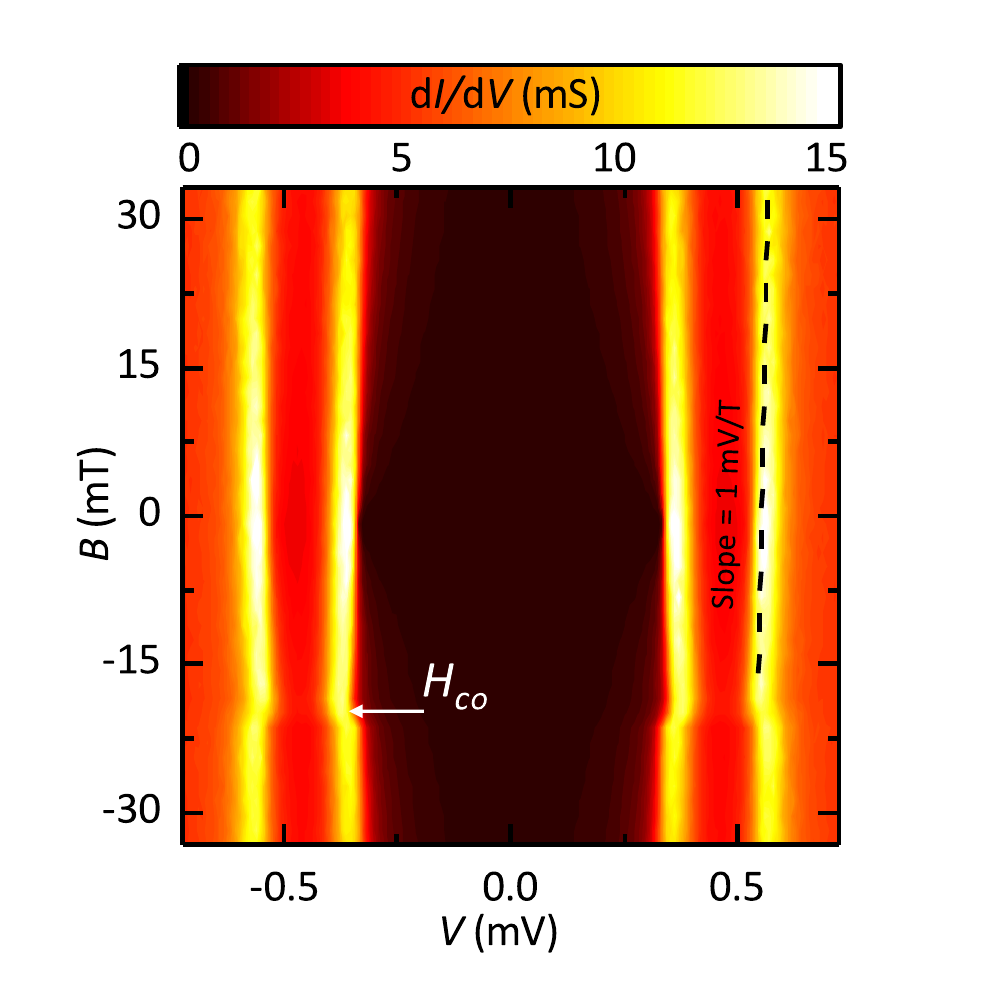}
\caption{Color plot of measured  differential conductance, $dI/dV$  of a  EuS/Al/Al$_2$O$_3$/Al  junction  as a function  of the  applied voltage and external magnetic field. $H_{co}$ denotes the  coercive field of the EuS layer when the magnetization switches. Figure adapted from the work by \textcite{strambini2017revealing}
 \label{fig:exp_eus_al}
}
\end{figure}

In Fig.~\ref{fig:exp_eus_al} we show an example of the measured  differential conductance  of an  EuS/Al/Al$_2$O$_3$/Al junction.   The Al layer adjacent  to the EuS has  a spin-split  density of states  that shows up as  the splitting peaks (bright  stripes  in the figure)  in $dI/dV$.  Even at zero applied magnetic field the  splitting is  nonzero.  The magnetization reversal of EuS  at  $H_c\approx -18.5$ mT manifests as a discontinuity  of the  conductance peaks \cite{strambini2017revealing}. As a first approach the DOS inferred from Fig. \ref{fig:exp_eus_al} can be well described by the expression (\ref{eq:split_Dos}). 

The advantage of using a FI instead of an external magnetic field 
 is that one avoids the depairing effects and all  complications  caused by  the need to apply magnetic fields in superconducting devices. 
Moreover,  because the electrons of the superconductor cannot propagate  into the FI,  superconducting properties are only modified by the induced spin-splitting field at the S/FI interface, and not by the leakage of Cooper pairs into the FI {as would happen in the case of metallic ferromagnets}. In addition, FIs can also be used as  spin-filter barriers \cite{Moodera_review}, in some cases with a very high spin-filtering efficiency, and therefore they play a crucial role for different 
applications as discussed below.

In Table  \ref{table:FIS} we  show  a list of FI/S combinations and the reported  induced  exchange splittings  and spin-filter efficiencies (barrier spin polarizations). 

\begin{table*}[t]
\caption{ Magnetic properties of different  ferromagnetic insulator-superconductor junctions used in experiments. Middle column shows the spin-filter efficiency characterized by the polarization $P=(G_\uparrow-G_\downarrow)/(G_\uparrow+G_\downarrow)$ of the FI barrier (red) with normal-state conductance $G_\sigma$ for spin $\sigma$. The exchange splittings measured in the superconductor (blue) are listed in the right column. The data is extracted from $^1$ \cite{PhysRevLett.56.1746}; $^2$ \cite{PhysRevLett.61.637}; $^3$ \cite{Hao:1990}; $^4$ \cite{PhysRevLett.70.853}; $^5$\cite{spin_filter_Blamire};$^6$ \cite{PhysRevB.92.180510}. Note that $\mu_B \cdot $1 T=58 $\mu$eV $\cong$ 670 mK. \label{table:FIS}}  
\begin{ruledtabular}
    \begin{tabular}{lll}
      Material Combination & Barrier polarization & Exchange Splitting (applied field)
      \\
      \hline
      \textcolor{red}{EuO}/\textcolor{blue}{Al}/AlO$_3$/Al $^1$ & no spin-filter barrier &1 T (0.1 T)-1.73 T(0.4 T)   \\
  Au/ \textcolor{red}{EuS}/\textcolor{blue}{Al} $^2$ & 0.8 &1.6 T (0 T)  \\
\textcolor{blue}{Al}/ \textcolor{red}{EuS}/Al $^3$&0.6-0.85 &1.9-2.6 T (0T) \\
Ag/ \textcolor{red}{EuSe}/\textcolor{blue}{Al} $^4$& $>$ 0.97 &none at zero field  \\
  \textcolor{red}{EuSe}/\textcolor{blue}{Al}/AlO$_3$/Ag $^4$& no spin-filter barrier & 4 T (0.6 T)  \\
  \textcolor{blue}{NbN}/ \textcolor{red}{GdN}/\textcolor{blue}{NbN} $^5$& 0.75 &   \\
     \textcolor{blue}{NbN}/ \textcolor{red}{GdN}/TiN $^6$& 0.97 & 1.4 T (0T)   \\
    \end{tabular}
\end{ruledtabular}
\end{table*}

\begin{figure}
\includegraphics[width=\columnwidth]{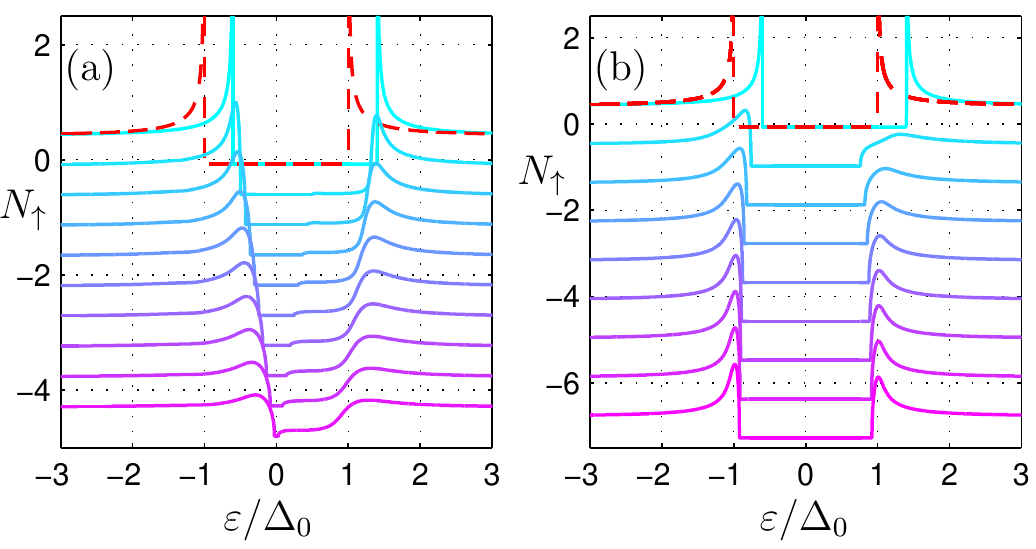}\caption{Calculated density of states of a thin superconducting film at $T\rightarrow0$. We only show the DOS for one
of the spin species, $N_\uparrow$.
Shown by dashed red lines is the DOS in the absence of relaxation $\tau_{sn}=1/(\tau_{sf}^{-1}+\tau_{so}^{-1}) =\infty$ and zero exchange field $h=0$ which corresponds to a gap $\Delta_0$. 
Other curves are plotted for $h=0.4\Delta_0$ and different spin relaxation rates. 
 (a) Spin-flip  relaxation $\beta=(\tau_{\rm so}-\tau_{\rm sf})/(\tau_{\rm so}+\tau_{\rm sf})=1$,
curves from top to bottom correspond to an increasing
$(\tau_{sn}\Delta_0)^{-1} $, varying equidistantly from $0$ by  $0.04$ steps.
(b) Spin-orbit relaxation $\beta=-1$, 
curves from top to bottom correspond to an increasing
$(\tau_{sn}\Delta_0)^{-1} $, varying equidistantly from $0$ by  steps of $3.4$.
For clarity the curves are shifted along the vertical axis.
\label{fig:Calculated-density-of}}
\end{figure}

{The paramagnetic effect, that leads to the spin-splitting,  is modified by spin relaxation and orbital depairing. In their absence} the superconductivity survives the spin-splitting field up to the Chandrasekhar-Clogston 
limit
\cite{PhysRevLett.9.266, Chandrasekhar:1962} $h=\Delta_0/\sqrt{2}${, where $\Delta_0$ is the order parameter at zero-field and  zero-temperature.} At this field the system experiences a first-order phase transition into the normal state when the order parameter changes abruptly from $\Delta_0$ to zero. This picture changes qualitatively due the presence of magnetic impurities and spin-orbit scattering. 
Even at $T=0$ and $h=0$ the spin-flip processes induced by magnetic impurities result in the pair breaking effect closing the energy gap \cite{Abrikosov1961}
 at $\tau_{sf}\Delta_0=3/4$ and suppresses  superconductivity completely at a certain  critical value of the spin-flip time 
 $\tau_{sf}$. For values of $\tau_{sf}$ larger than the critical one the phase transition switches from the first to the second order at \cite{bruno1973-mfs}  $\tau_{sf}\Delta_0=0.461$ and the gapless state appears at a certain value of $h(\tau_{sf})$ (see Fig.~\ref{fig:Calculated-density-of}a) .
 
Contrary to the spin-flip processes, the spin-orbit scattering alone does not have any effect on the superconducting state. However, in combination with $h\neq 0$ it tends to smear out the spin-splitted DOS singularities  provided the spin-orbit  relaxation time, $\tau_{so}$, is not very short (see Fig.~\ref{fig:Calculated-density-of}b). At short relaxation times
{ $\tau_{so} \ll \Gamma/\Delta^2$, where $\Gamma$ is the depairing parameter \cite{dynes84} } the effect of 
  spin splitting is eliminated and the usual BCS  density of states is recovered (see Fig.~\ref{fig:Calculated-density-of}b). Therefore in this case the critical  spin-splitting field is strongly increased  above the Chandrasekhar-Clogston limit\cite{bruno1973-mfs}. 
  
 Besides broadening of the DOS singularities, the spin-orbit and spin-flip relaxation processes  have an important effect on the paramagnetic spin susceptibility of the superconductor as it becomes non-vanishing even in the zero-temperature limit \cite{yosida1958paramagnetic, abrikosov1961problem,bruno1973-mfs}. 
 The static spin susceptibility characterizes the paramagnetic response of the superconductor to an external magnetic field. In a usual normal metal the Zeeman field produces the same magnetization as a spin-dependent chemical potential shift  
 $\delta \mu$ of the same magnitude when the distribution functions in different spin subbands are given by 
 $f_{\uparrow}(E) = f_0(E+\delta\mu)$ and $f_{\downarrow}(E) = f_0(E-\delta\mu)$. This is different in superconductors  where the paramagnetic susceptibility is determined by both the spin-polarized quasiparticles and the emergent spin-triplet superconducting correlations \citep{Abrikosov1962, 
abrikosov1961problem}. On the other hand, the non-equilibrium spin modes as systematically described in Sec.~\ref{sec:modes} are determined only by the quasiparticle contribution.

In the next sections we review the transport properties of diffusive hybrid structures with  spin-split superconductors by contrasting existing theories and experiments.  For this sake, in the next section  we briefly introduce  the quasiclassical Green's function  formalism for superconductors in the presence of spin-dependent fields and spin-polarised interfaces. {It is in our opinion} the most suitable formalism for the description of diffusive hybrid structures. 

\subsection{Brief overview of  the quasiclassical theory of diffusive superconductors\label{sec:Usadel}}

Quasiclassical Keldysh Green's function technique is a useful and well-established way to describe transport and nonequilibrium properties of good metals, where the relevant physical length scales affecting different observables are long compared to the Fermi wave length $\lambda_F$, and where in particular disorder plays a major role. Several reviews explain this technique for various applications  \cite{belzig1999quasiclassical,Bergeret2005}. Here we just outline the main features relevant for spin-split superconductors. Briefly, the Keldysh Green's functions (GFs), $\check G(\bf{r},\bf{r'};t,t')$ are two-point correlation functions which depend on two coordinates and two times. Here the ``check'' $\check G$ denotes GFs that live in a structure formed by the direct product of Keldysh, spin and Nambu spaces.  The  equation of motion for $\check G$ can be {written as} a kinetic-like equation for the Wigner transformed GF, $\check G(\bf{R},\bf{p})$, where $\bf{R}$ and $\bf{p}$ are the center of mass coordinate and $\bf{p}$ the momentum after Fourier transformation with respect to the relative component.  A significant simplification can be done in the case of metals by noticing that the Green's functions are peaked at the Fermi level. This allows for an   integration of the equations over the quasiparticle energy, {related to} the magnitude of $\bf{p}$. This procedure leads to the quasiclassical GFs, $\check g (\bf{R},\bf{n})$, which only depend on the direction of the momentum at the Fermi level and on two times in the case of non-stationary problems, or only on a single energy $\varepsilon$ in the stationary case. These functions obey the \textcite{Eilenberger1968} equation. One of the advantages of using the quasiclassical GFs is that in the normal state, the spectral part is trivial, {\it i.e.}, the retarded and advanced GFs are energy independent. All transport information of the normal metal is encoded in the quasiclassical Wigner  distribution function $\hat f (\bf{R},\bf{n})$ and quasiclassical equation for it resembles the classical Boltzmann equation \cite{Langenberg1986}.

In contrast, the superconducting case distinguishes itself by a non-trivial spectrum, and therefore requires taking into account the full Keldysh structure of the GFs, {\it i.e.}
\begin{equation}
{\check g}=\left(\begin{array}{cc}
{\hat g^{R}} & {\hat g^{K}}\\
0 & {\hat g^{A}}
\end{array}\right)\;. \label{eq:gmatrix}
\end{equation} 
This GF satisfies the normalization condition\cite{Eilenberger1968}
\begin{equation}
\check{g}^{2}=\check{1}\;.\label{normalization}
\end{equation}

In the  diffusive limit the elastic mean free path $l$  due to scattering at non-magnetic impurities  is much smaller  than any other length involved in the problem except $\lambda_F$. 
Within this limit the Eilenberger equation can be reduced to a diffusive-like equation,  in the same way as the Boltzmann equation is simplified in the  diffusive limit.  This quasiclassical diffusion equation for superconductors is the \textcite{usadel.1970} equation (we set $\hbar=k_B=1$)
\begin{equation}
 D\nabla\cdot(\check{g}\nabla\check{g})+[i \varepsilon\tau_{3}-i\vec{h}\cdot\vec{\sigma}\tau_{3}-\check{\Delta}-
 \check\Sigma,\check{g}]=0.\label{eq:Usadel}
 \end{equation}
 Here $D$ is the diffusion coefficient, 
 $\check{g}({\bf r},\varepsilon)$ is the isotropic (momentum independent)  quasiclassical GF,  $\bm{h}$  the spin-splitting field either generated
by an external field or by the magnetic proximity effect in a FI/S junction, and $\check{\Delta}=\Delta e^{i\varphi \tau_3}\tau_1$ depends on the superconducting  order
parameter $\Delta$ that has to be determined  self-consistently. Here $\tau_i$ and $\sigma_i$ are Pauli spin matrices in Nambu and spin space, respectively. The self-energy $\check{\Sigma}$ in Eq.~(\ref{eq:Usadel}) describes different scattering processes,  such as elastic spin-flip or spin-orbit scattering, $\check \Sigma_{\rm el}$ and inelastic electron-phonon and electron-electron scattering, $\check \Sigma_{\rm in}$.

Equation \eqref{eq:Usadel} is central in the description of diffusive superconducting structures.  Whereas the spectral properties can be obtained by solving the retarded {(R)} and advanced {(A)} components of this equation, nonequilibrium properties  are described by the kinetic equation obtained by taking  the Keldysh (K) component of  Eq. (\ref{eq:Usadel}). This can be compactly written as
\begin{equation}
\nabla_k j_{kb}^a = H^{ab}+R^{ab}+I_{\rm coll}^{ab},
\label{eq:kineticeq}
\end{equation}
where we introduce the spectral current tensor $j_{kb}^{a}$, 
\begin{equation} 
j_{kb}^{a}=\frac{1}{8}{\rm Tr}\tau_{b}\sigma_{a}(\check{g}\nabla_{k}\check{g})^{K}\;.
\label{eq:general_current}
\end{equation}
The different  current density components (charge, spin, energy, spin-energy)  can be obtained from  Eq~(\ref{eq:general_current}). For example, the charge current density reads
\begin{equation} 
  J_{k} = \frac{\sigma_{N}}{2e}\int_{-\infty}^\infty d\varepsilon\,j_{k3}^0,
  \label{eq:charge_current}
\end{equation}
Here $\sigma_N=e^2 \nu_F D$ and $\nu_F$ are the normal-state conductivity and density of states at the Fermi level respectively. In Eq.~(\ref{eq:kineticeq}) the term $H^{ab}={\rm Tr} \tau_b \sigma_a [-i{\mathbf h} \cdot \bm{\sigma} \tau_3,\hat g^K]/8$ describes the Hanle precession of spin caused by the exchange field, and $R^{ab}={\rm Tr} \tau_b \sigma_a [\hat\Delta,\hat g^K]/8$ the conversion between quasiparticles and the superconducting condensate. 
Finally  $I_{\rm coll}^{ab}={\rm Tr} \tau_b \sigma_a [\check \Sigma,\check g]^K/8$ in Eq.~(\ref{eq:kineticeq})  is the  collision integral describing the different  scattering process with self-energy $\check \Sigma$. We discuss next different scattering processes.

\emph{Elastic self-energy terms.} 
We consider  elastic contributions to $\check{\Sigma}_{el}$  due to scattering at impurities
with spin-orbit coupling (relaxation time $\tau_{so}$) and
the spin flips at magnetic impurities ($\tau_{sf}$) \cite{maki1966effect}.  
{Within the Born approximation, they read
$\check{\Sigma}_{so}=\vec{\sigma}\cdot\check{g}\vec{\sigma}/(8\tau_{so})$,
$\check{\Sigma}_{sf}=\vec{\sigma}\cdot\tau_{3}\check{g}\tau_{3}\vec{\sigma}/(8\tau_{sf})$.}
In the normal state they contribute to the energy-independent total spin-relaxation time 
$\tau_{\rm sn}^{-1}=\tau_{\rm so}^{-1}+\tau_{\rm sf}^{-1}$. In contrast, in the superconducting case the spin-relaxation time and length acquire  energy dependence, which is different for the spin-orbit and spin-flip scattering \cite{maki1966effect,morten2004spin,morten2005spin}. 
Therefore it is convenient to describe the relative strength of these two scattering mechanisms in terms of the parameter
$\beta=(\tau_{\rm so}-\tau_{\rm sf})/(\tau_{\rm so}+\tau_{\rm sf})$. 
In  diffusive superconducting thin films one can also describe the depairing effect of an in-plane magnetic field with a  self-energy term $\check{\Sigma}_{orb}=\tau_{3}\check{g}\tau_{3}/\tau_{orb}$ characterized by the orbital depairing time $\tau_{orb}$ \cite{deGennes:566105,anthore2003}. This term  also contributes to charge imbalance relaxation \cite{schmid1975,nielsen1982pair}.

The parameters $\tau_{\rm sn}^{-1}$ and $\beta$ are material specific. For example, in Al films, the reported values from a set of spin injection experiments are $\tau_{\rm sn} \approx 100$ ps \cite{jedema:713,poli2008spin} and $\beta \approx 0.5$ indicating the dominance of spin-flip relaxation over spin-orbit scattering, whereas the reported value of $\tau_{sn}$ in Nb is only 0.2 ps, and is strongly dominated by spin-orbit scattering \cite{wakamura2014spin}. They affect both the spectrum of a bulk superconductor (see Fig.~\ref{fig:Calculated-density-of}) and the spin relaxation as described in Sec.~\ref{sec:nl_detection}. 

\emph{Inelastic self-energies.} The relevant inelastic processes entering the self-energy in Eq.~\eqref{eq:Usadel}, are the particle--phonon and particle--particle
collisions.
These
processes do not conserve the energies of colliding quasiparticles,
but conserve the total spin.  

The coupling between quasiparticles and phonons limits some of the effects
discussed in the following sections. Due to the energy dependence of
the phonon density of states, this coupling decreases
rapidly towards low temperatures, and eventually phonons decouple from
electrons, and the main heat relaxation occurs via other processes
such as quasiparticle diffusion. Superconductivity
modifies the electron-phonon heat conduction
\cite{eliashberg1972-iec,kopnin2001-ton,kaplan1976}, as also the
electronic spectrum is energy dependent, and is affected by the
spin splitting \cite{grimaldi1997,virtanen2016stimulated}.

Particle-particle collisions in superconductors and superfluids are
discussed by \textcite{eliashberg1972-iec,serene1983-qat,kopnin2001-ton},
although mainly within contact interaction models disregarding
screening effects \cite{narozhny1999,feigelman2000,kamenev2009}. The
collision integrals can have spin structure also in the normal state
\cite{Dimitrova2008,chtchelkatchev2008-ers}.

The far-from-equilibrium results discussed in Sec.~\ref{spininjection}
disregard the particle-particle collisions, as the simpler theory already
describes effects not very far from the measured ones. On the other
hand, Sec.~\ref{thermoel} mostly concentrates on the quasiequilibrium
limit, where also spin accumulation is lost due to a strong spin
relaxation.

\emph{Hybrid interfaces}. In subsequent sections we apply the kinetic equation, Eq.~(\ref{eq:kineticeq}),  in  different situations. For the description of transport in  hybrid structures,  we need in addition 
a description  of  interfaces between different materials in the form of boundary conditions.
Such interfaces usually are described by sharp changes of the potential and material parameters 
over atomic distances, and thus cannot be included 
{directly} in the quasiclassical equations which describe properties over distances much larger than $\lambda_F$.   
 The  description of hybrid interfaces  requires  then the derivation of suitable 
boundary conditions, first done in the quasiclassical approach by \textcite{zaitsev1984quasiclassical}.

Boundary conditions for the Usadel equation trace back to the work of
\textcite{Kupriyanov1988}. These boundary conditions are
applicable for non-magnetic N-N, S-S and S-N interfaces with low
transmissivity \cite{PhysRevB.55.6015}. Later {\textcite{nazarov.1999}} generalized
these boundary conditions for an arbitrary interface transparency.

\textcite{PhysRevB.38.8823} derived
the boundary condition for an interface between a superconductor and
a ferromagnetic insulator and introduced the concept of the spin-mixing
angle, which describes the spin-dependent phase shifts acquired by the  electrons after 
being   scattered
at the FI/S interface. Later \textcite{PhysRevB.70.134510,PhysRevB.80.184511} extended these boundary
conditions to magnetic metallic structures, such as F-S or S-F-S systems,
though with low polarization. Boundary
conditions for large polarization and low transmission have been presented
by \textcite{Bergeret2012,Machon2013}. General boundary conditions for arbitrary spin polarization  and 
transmission have been extensively discussed by  \textcite{eschrig2015general}.

Here we mainly deal with low transmissive 
barriers between a mesoscopic   superconductor and normal and magnetic
leads and use the description presented by \textcite{Bergeret2012}. {In this description, the component of the spectral current density multiplied by $\sigma_N$ perpendicular to the interface is continuous across it, and given by
\begin{align}
  \sigma_N
  j_{\perp,b}^a = -\frac{1}{8 e R_{\square}} 
  {\rm Tr}\tau_{b}\sigma_{a}\left[\hat{\Gamma}\check{g}_{2}\hat\Gamma^{\dagger},\check{g}\right]^{K}
  \label{eq:currents_at_interface}
  \,,
\end{align}
where $R_{\square}$ is the spin-averaged barrier resistance per unit area,
and the spin-dependent transmission is characterized by the tunneling matrix  $\hat{\Gamma}=t\tau_{3}+u\sigma_{3}$, assuming polarization in the $z$-direction. The normalized transparencies satisfy $t^{2}+u^{2}=1$ and are determined from the interface polarization $|P|\le1$ via $2ut=P$. The Green's function $\check{g}_{2}$ in the r.h.s of Eq. (\ref{eq:currents_at_interface}) corresponds to the opposite side of the junction.}

\section{Nonequilibrium modes  in spin-split superconductors}\label{sec:modes}

The out-of-equilibrium state in superconducting systems is characterized by the presence of nonequilibrium modes associated with the different electronic degrees of freedom. 
For example, injection of an electric current
from a normal  electrode into a  superconductor generates a charge imbalance mode
\cite{PhysRevLett.28.1363,PhysRevB.6.1747,PhysRevLett.28.1366,yagi2006charge,hubler2010charge} that diffuses into the S region. 
This nonequilibrium mode reflects
an imbalance of the  quasiparticle population between the electron-like and hole-like
spectrum branches. The charge imbalance measurements made in the 1970s were to our knowledge the first to study such nonequilibrium modes in non-local multiterminal settings. This technique was later adapted to spintronics, to study the nonequilibrium spin accumulation induced by spin-polarized electrodes \cite{PhysRevLett.55.1790}.

\begin{figure*}
  \includegraphics[scale=0.3]{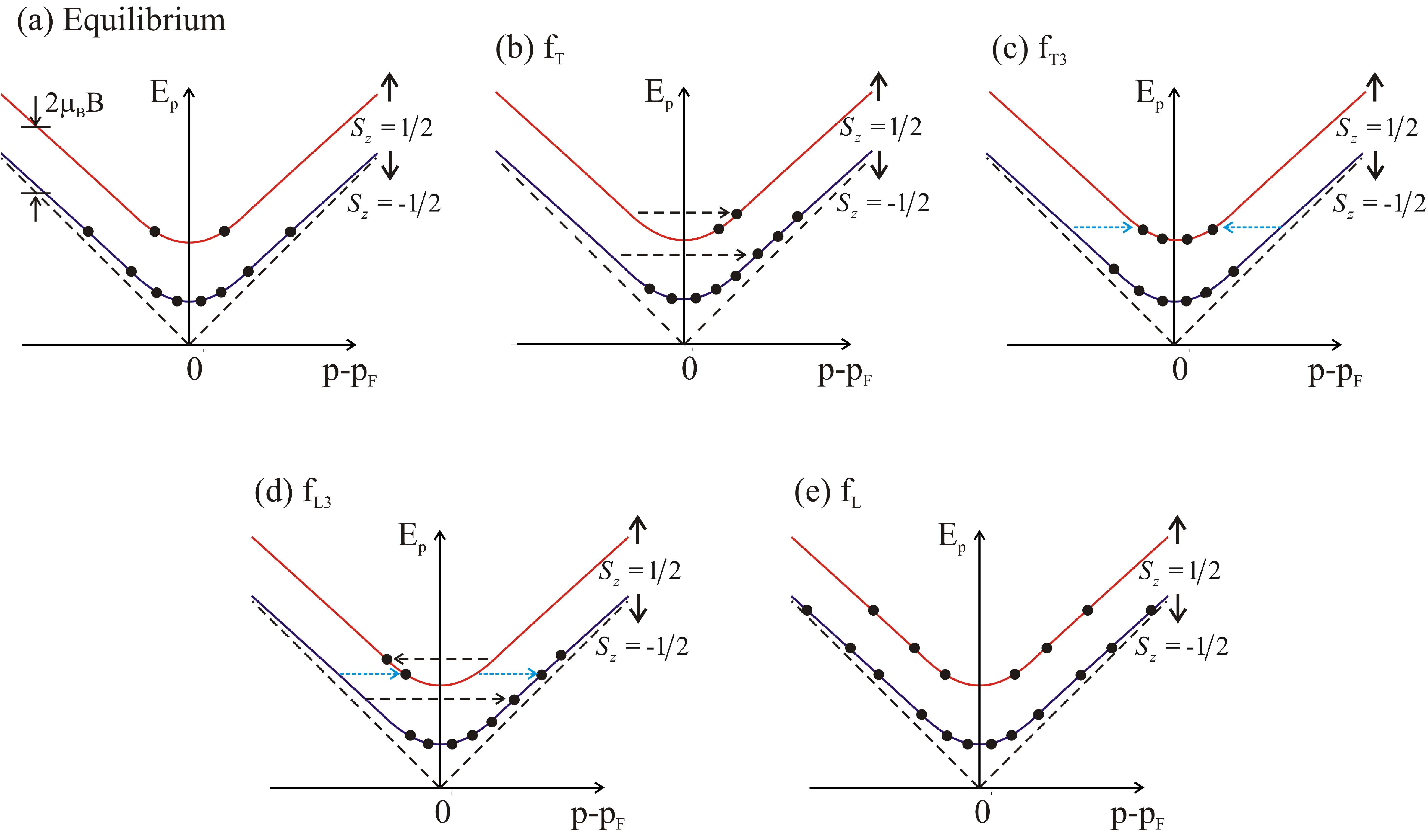} 
  \caption{\label{fig:DistributionFunction}
    Schematic illustration of the quasiparticle distribution function components  
    in a superconductor with spin splitting $2\mu_B B$. 
    The occupied states are  represented by filled circles. 
    (a) Equilibrium distribution, (b) charge imbalance $f_{T}$,
    (c) spin imbalance $f_{T3}$, (d) spin energy imbalance $f_{L3}$, and (e) energy imbalance $f_{L}$. The dashed and dotted arrows show elastic processes which 
    lead to the  formation --- and the reverse processes to the relaxation --- 
    of a particular nonequilibrium mode. In (c,d) the dashed black lines
    show particle-hole branch transitions while the dotted blue lines
    correspond to the spin-flip processes. 
    }
\end{figure*}

Schematically, nonequilibrium modes can be
represented in terms of the electron/hole branches in the  spectrum of the superconductor \cite{TinkhamBook}, as illustrated in Fig.~\ref{fig:DistributionFunction}. 
For example the charge mode can be understood as the imbalance between the electron and hole branches (Fig.~\ref{fig:DistributionFunction}b).
In the absence of  spin-dependent fields there is one more nonequilibrium
mode: the energy imbalance mode (Fig.~\ref{fig:DistributionFunction}e) .
It describes 
the excess energy  stemming from an equal  change in the quasiparticle populations of the electron-like and hole-like branches.
This energy mode affects charge transport properties indirectly via the self-consistency equation for $\Delta$.  
This mechanism explains, for example, the enhancement of the superconducting transition
temperature in the presence of a microwave
field \cite{Eliashberg_enhacement,Klapwijk_enhancement}.  

In this section we generalize the  description in terms of  nonequilibrium  modes  to account for superconductors with spin-split density of states.   The spin splitting {(energy difference $2h=2\mu_B B$ between the black and red dispersion curves in Fig.~\ref{fig:DistributionFunction} for spin up/down quasiparticles)}   gives rise to four distinct quasiparticle
branches, electron/hole and spin up/down. 
These four nonequilibrium modes and their coupling  are at the basis of the main effects  discussed in this review. 

\subsection{Description of nonequilibrium modes in superconductors with spin splitting} \label{SubSec:Modes}

At this point {we} combine the pictorial description of the nonequilibrium modes (Fig.~\ref{fig:DistributionFunction}) with the quasiclassical formalism introduced in Sec.~\ref{sec:Usadel} and {in particular,} the Usadel equation. For a description of non-equilibrium properties we need  to  consider the  Keldysh component $\hat g^K$ of the quasiclassical GF [Eq.~(\ref{eq:gmatrix})]. For clarity  we first consider a unique spin polarization direction parallel to the $z$-axis.  From  the normalization condition, Eq.~(\ref{normalization}), $\hat g^K$ can be expressed in terms of the retarded and advanced components and  the  generalized matrix distribution function $\hat f$ \cite{Langenberg1986}
\begin{equation}
\hat{g}^{K}=\hat{g}^{R}\hat{f}-\hat{f}\hat{g}^{A}\;.\label{eq:K_param}
\end{equation}
In the case of only one spin polarization axis,  the   4$\times$4 matrix distribution function  $\hat{f}$ can be written as the sum of four components\footnote{Here we assume a unique spin polarization direction. In the most general case the distribution function has all spin components $
 \hat{f}=f_{L}\hat{1}+f_{T}\tau_{3}+\sum_{j}(f_{Tj}\sigma_{j}+f_{Lj}\sigma_{j}\tau_{3})\label{eq:distribution_function}$.}
 
\begin{equation}
\hat{f}=f_{L}\hat{1}+f_{T}\tau_{3}+(f_{T3}\sigma_{3}+f_{L3}\sigma_{3}\tau_{3})\;.\label{eq:distribution_function0}
\end{equation}

For historical reasons  we use the labeling introduced by  \textcite{schmid1975}, generalized for the spin-dependent case.
The $L$-labeled
functions describe longitudinal modes, the (spin) energy degrees
of freedom, and are antisymmetric in energy with respect to the Fermi
level, $\varepsilon=0$. The $T$-labeled functions describe transverse
modes and are symmetric in energy. In equilibrium, the
distribution function is proportional to the unit matrix in Nambu
and spin space, and given by 
\begin{equation}
\hat{f}_{eq}(\varepsilon)=(1-2n_F)\hat 1=\tanh(\varepsilon/2T)\hat 1\; .\label{eq:feq} 
\end{equation}
We can now turn to the pictorial description of Fig.~\ref{fig:DistributionFunction} and associate   each component of $\check f$ in Eq.~(\ref{eq:distribution_function0})  with a nonequilibrium mode as discussed  next. 
 
As shown  in Figs.~\ref{fig:DistributionFunction}(b)--(e), two of these modes have electron-hole
branch imbalance, $f_T$ and  $f_{L3}$, while $f_{T3}$ and $f_L$ are particle-hole symmetric.  The filled circles in Fig.~\ref{fig:DistributionFunction} represent the occupied states.  
As a reference, panel (a)  corresponds to the equilibrium distribution function  $\hat f = f_L^0 \hat 1=\tanh(\varepsilon/2T) \hat 1$.  In order to excite the nonequilibrium modes, $f_T$, $f_{T3}$ and $f_{L3}$, one only needs to move 
the populated states (filled circles) between the different spectral branches in an elastic process,
 i.e., between equal-energy states (marked by horizontal dashed arrows). 
These modes can also relax back to equilibrium due to elastic scattering processes. 
The relaxation mechanisms depend on intrinsic material properties,
degree and type of disorder, and also on the superconducting spectrum,
and are discussed in more detail below.

The last nonequilibrium mode, the deviation of $f_L$ from $f_L^0$, is characterized by a change in the total quasiparticle number and energy content, corresponding to an increase or decrease of the effective temperature. It can be excited by increasing the number of occupied states to higher energies,
and its relaxation requires inelastic processes.

In the absence of spin splitting, the charge imbalance is determined by
$f_T$, and the energy imbalance by $f_L$.
The spin splitting changes the system properties, mixing the coupling between spin-dependent modes and physical observables [see Eqs.~(\ref{Eq:ChPot0}) and (\ref{Eq:ChPotZ}) below]. Qualitatively, the outcome can be seen by counting the
number of occupied states on the different branches in Fig.~\ref{fig:DistributionFunction}. 
For example, the charge imbalance $\mu$ is determined by the difference between the number of occupied states
in the electron and hole branches. Both
 $f_T$ and $f_{L3}$ components contribute to it, as seen in
 Figs.~\ref{fig:DistributionFunction}b and d. 

On the other hand, a nonzero spin accumulation $\mu_z$ can be induced by exciting the  modes 
$f_{T3}$ or $f_L$ [Figs.~\ref{fig:DistributionFunction}(c),(e)]. 
These two contributions to the total spin accumulation have important differences:
The mode $f_{T3}$ contributes to spin imbalance also  
in the absence of spin splitting.  Spin imbalance in this mode can be induced for example
by a spin-polarized injection from a
ferromagnetic electrode, in both the normal and the superconducting state. 
The relaxation of the spin accumulation created in this way is 
determined by elastic scattering processes.  The second mechanism 
of inducing spin accumulation is by exciting  the longitudinal mode  $f_L$,
in the presence of spin splitting [Fig.~\ref{fig:DistributionFunction}(e)].
Since  energy-conserving transitions do not result in the relaxation of the $f_L$ mode,
this component of the spin imbalance is not suppressed by elastic scattering.
In other words, its relaxation can be only provided by inelastic
processes, e.g., electron-phonon and electron-electron scattering. 
This result, obtained here on a phenomenological level, is crucial in understanding 
the long-range spin  signal observed in superconductors, for example by \textcite{Hubler2012a} and discussed in the next sections.

\subsection{{Accumulations}
  in terms of the non-equilibrium modes \label{sec:noneq_quasiclas}}

{Quantitatively, we define the charge and spin accumulations} based on the  Keldysh
component of the GF, Eq.~\eqref{eq:K_param},
\begin{eqnarray}
\mu({\bf r},t) & = & -\int_{-\infty}^{\infty} \frac{d\varepsilon}{16}  \Tr \hat g^{K}(\varepsilon,{\bf r},t)\label{eq:def_mu}\\
\mu_{sa}({\bf r},t) & = & \int_{-\infty}^{\infty} \frac{d\varepsilon}{16} \Tr \tau_{3}\sigma_{a}[ \hat g_{\rm eq}^{K}(\epsilon,\mathbf{r},t)- \hat  g^{K}(\varepsilon,\mathbf{r},t)]\; , \label{eq:def_mus}
\end{eqnarray}
whereas the local energy and spin-energy accumulations are given by
\begin{eqnarray} \label{Eq:HeatDens}
  q({\bf r},t) & = & \int_{-\infty}^{\infty} \frac{d\varepsilon}{16}  \varepsilon \Tr \tau_3 [\hat  g_{\rm eq}^{K}(\varepsilon,{\bf r},t)- \hat  g^{K}(\varepsilon,\mathbf{r},t)]\label{eq:def_q}
  \\  \label{Eq:SpinEnergyDens}
  q_{sa}({\bf r},t) & = &
  \int_{-\infty}^{\infty} \frac{d\varepsilon}{16} \varepsilon \Tr\sigma_{a}[ \hat  g_{\rm eq}^{K}(\varepsilon,\mathbf{r},t)- \hat  g^{K}(\varepsilon,\mathbf{r},t)]. \label{eq:def_qs}\;
\end{eqnarray}
Above, $a=1,2,3$ denotes the polarization direction of the
nonequilibrium spins and energy is counted with respect to the
potential $\mu_S$ of the superconducting condensate (see below).

In terms of the distribution functions, the charge and spin accumulations read {(we assume  magnetization in $z$-direction)}
\begin{align} \label{Eq:ChPot0}
 & \mu = -\frac{1}{2}\int_{-\infty}^{\infty} d\varepsilon ( N_+ f_T+ N_-  f_{L3}) \\\label{Eq:ChPotZ}
 & \mu_z = - \frac{1}{2}\int_{-\infty}^{\infty} d\varepsilon [ N_+ f_{T3}+ N_-(f_{L}-f_{\rm eq})],
\end{align}
where $N_+ = N_\uparrow + N_\downarrow$ is the total density of states
(DOS), $N_- = N_\uparrow - N_\downarrow $ is the DOS difference
between the spin subbands, and $f_{\rm eq}(\varepsilon) =
\tanh(\varepsilon/2T)$ is the equilibrium distribution function. 
{Similarly for  (\ref{Eq:HeatDens},\ref{Eq:SpinEnergyDens}) we get 
\begin{align} \label{Eq:HeatDensQuasi}
 & q =  \frac{1}{2}\int_{-\infty}^{\infty} d\varepsilon \varepsilon [ N_+ (f_L-f_{eq}) + N_-f_{T3}]
\\ \label{Eq:SpinEnergyDensQuasi}
 & q_{sa} =  \frac{1}{2}\int_{-\infty}^{\infty} d\varepsilon \varepsilon [ N_- (f_{L}-f_{\rm eq}) + N_+f_{T3}].
\end{align}
}
{ All these quantities, Eqs. (\ref{eq:def_mu}-\ref{eq:def_qs}) are  directly related to experimental observables.
}{
The charge imbalance $\mu$ characterizes the potential of the quasiparticles
in the superconductor \cite{artemenko1979-efc}.
In nonequilibrium situations, $\mu$ can differ from the condensate potential
$\mu_S$.
In the problems discussed in this Colloquium, $\Delta$ can be chosen time-independent and $\mu_S=0$.
The charge density depends on $\mu$ via $\rho=-\nu_Fe^2\phi-e\nu_F\mu$
where $\phi$ is the electrostatic scalar potential \cite{kopnin2001-ton}.
In metals, local charge neutrality is maintained on length scales large compared
to the Thomas--Fermi screening length, so that $-e\phi=\mu$ and charge imbalance is associated with static electric fields.

In the quasiclassical formulation used here, electrochemical potential differences
appear explicitly in energy shifts in the boundary conditions for the
distribution functions \cite{belzig1999quasiclassical}. The Fermi
distribution at potential $V$ corresponds to
\begin{align}
  f_{\mathrm{eq},L(T)}(E)=\frac{1}{2}[\tanh\bigl(\frac{E+eV}{2T}\bigr) +(-)
  \tanh\bigl(\frac{E-eV}{2T}\bigr)] \,.
\end{align}
For superconductor at equilibrium, $V=0$ in this description.
However, $V=\phi\ne0$ can describe voltage-biased normal ($\Delta=0$) reservoirs.
}

Spin accumulation is a standard observable in spintronics \cite{PhysRevLett.55.1790,jedema:713}.
The local energy accumulation is typically measured via electron thermometry \cite{giazotto2006}. The spin-energy accumulation was measured recently in normal-state nanopillar spin valves \cite{dejene2013}. To our knowledge this quantity has not been directly studied experimentally in superconducting systems.

In the normal state the spectrum is trivial,  $g^{R(A)}=\pm \tau_3\sigma_0$. Thus, according to Eq.~(\ref{eq:K_param}), the Keldysh component  is simply proportional to the distribution function.  In other words, the different {modes decouple} in Eqs. (\ref{eq:def_mu}-\ref{eq:def_qs}). 
Moreover, in the normal state it is unnecessary to separate between transverse and longitudinal modes, and rather consider the spin-dependent full distribution function $f_j(E)=[1-f_{Lj}(E)-f_{Tj}(E)]/2$.   Solutions of the kinetic equation in the normal state  are discussed for example by \textcite{Brataas:2006en}.

In the superconducting case the situation is more  complex. First,  the spectrum is strongly energy dependent around the Fermi level and the spectral GFs 
have a non-trivial structure in spin space. 
Components  proportional to the unit matrix in spin space   describes the BCS singlet GFs, whereas 
terms proportional to the Pauli matrices
 $\sigma_{j}$, $j=1,2,3$, describe the triplet state \cite{bergeret2001long,Bergeret2005}.
Second, due to this energy dependence and non-trivial spin structure, the  spectral functions  enter
(\ref{eq:def_mu}-\ref{eq:def_qs}) and lead to a coupling between the different non-equilibrium modes that in turns couple all electronic  degrees of freedom, as  discussed next.

\section{Spin injection and diffusion in superconductors\label{spininjection}}

Non-equilibrium modes can be experimentally studied by means of  non-local transport measurements. In this section we review experiments on charge and spin injection in superconductors, and {apply the kinetic equation approach described in the} previous sections to describe different experimental situations.

\subsection{Detection of spin and charge imbalance: Non-local transport measurements\label{sec:nl_detection}}
 \begin{figure}
\protect\includegraphics[width=\columnwidth]{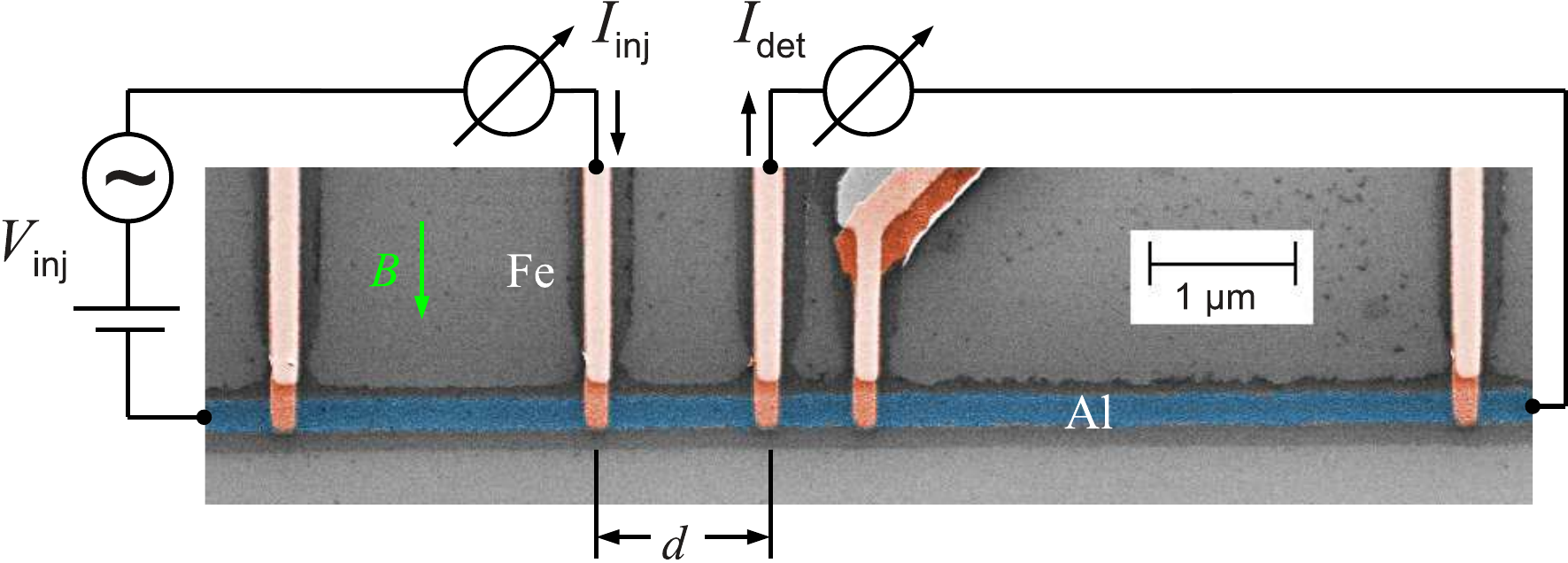}
\caption{Scanning electron microscopy image of the lateral structure used by \textcite{Hubler2012a}.
  From \textcite{Hubler2012a}.
}
 \label{fig:experiemnt_beckmann}
\end{figure}

Studies of the nonequilibrium modes started   with the pioneering experiment of  \textcite{PhysRevLett.28.1363}, who  realized  a  way of detecting the charge imbalance in a superconductor.  
The main idea of this experiment 
is to inject 
a current from a normal metal (injector) into a superconductor.  
This current  creates  a charge imbalance that  {corresponds to} a shift of the chemical potential of the quasiparticles with respect to the one of the condensate.  This shift of the chemical potential can be detected by a second electrode (detector) that probes the voltage between the superconductor and the detector.

More recent experiments used the same non-local measurement to 
explore the charge, energy and spin modes in mesoscopic superconducting  lateral structures \cite{beckmann2004evidence,hubler2010charge,wolf2014charge,PhysRevB.87.024517,Quay2013,poli2008spin}. A scanning electron microscopy image of such a lateral structure is shown in Fig.~\ref{fig:experiemnt_beckmann}.  
A detailed overview of the experiments on charge and energy imbalance can be found in the recent topical reviews  by \textcite{beckmann2016spin,quay2017up}.

Whereas the charge and energy modes were known for a long time, it was first   in the 1990s that theorists predicted that electronic  charge and spin degrees of freedom can be separated in a superconductor
\cite{Kivelson1990,PhysRevB.52.3632}.  First experiments  on F-S-F layered structures,  however,  did not show any evidence of
such a spin-charge separation \cite{Johnson1994,gu2002direct} and the different relaxation times  for  spin and charge accumulation in superconductors remained an open question.

First clear insight into  the separation of the spin and charge modes  was obtained in experiments using    lateral  nanostructures  with ferromagnetic injectors and detectors \cite{beckmann2004evidence,PhysRevB.71.144513,cadden2007charge,Hubler2012a,Kolenda2016,PhysRevB.87.024517,Wolf2014,poli2008spin,Quay2013,yang2010extremely}.  First theoretical works on spin injection into mesoscopic superconductors \cite{morten2004spin,morten2005spin} showed 
that the  spin-relaxation length   in the superconducting state 
 strongly  depends on the energy of the injected quasiparticles and on the spin relaxation mechanism. In particular, for a dominating spin-orbit scattering, superconductivity suppresses the
 spin relaxation {rate $\tau_s^{-1}$}, which can be  qualitatively  understood as the decrease in the cross section of the quasiparticle momentum scattering at the energies near the gap edge $\varepsilon\sim \Delta$.  The suppression of $\tau_{s}^{-1}$ is however compensated by the 
decrease in the quasiparticle group velocity 
$v_g\sim{}v_F\sqrt{1-|\Delta|^2/\varepsilon^2}$ so that the spin relaxation length $\lambda_{so} \sim v_g \tau_{s} $ remains almost unchanged in the superconducting state. On the contrary, if the spin-flip mechanism dominates, the spin relaxation is not related to the momentum scattering because the interaction with magnetic impurities does not depend on the propagation direction and the quasiparticle spin does not depend on energy.  This results in an increase of  $\tau_{s}^{-1}$ which is equivalent to a decrease of the spin-relaxation length in the superconducting state. Although  these works provided an explanation to some experiments,  two important  features  observed in subsequent works could not be explained in terms of that  theory: 
First, the spin accumulation {was} detected at distances from the
injector much larger than the spin-relaxation length measured in the normal state \cite{Quay2013,Hubler2012a,PhysRevB.87.024517}. Second,
 an unexpected  spin accumulation was observed even if the current was injected from a non-magnetic electrode \cite{PhysRevB.87.024517}.  { In order to explain these two observations one needs to take into account the spin splitting in the superconductor. }

\subsection{Non-local conductance measurements in spin-split superconductors}

\begin{figure}
 \centerline{\includegraphics[width=\columnwidth]{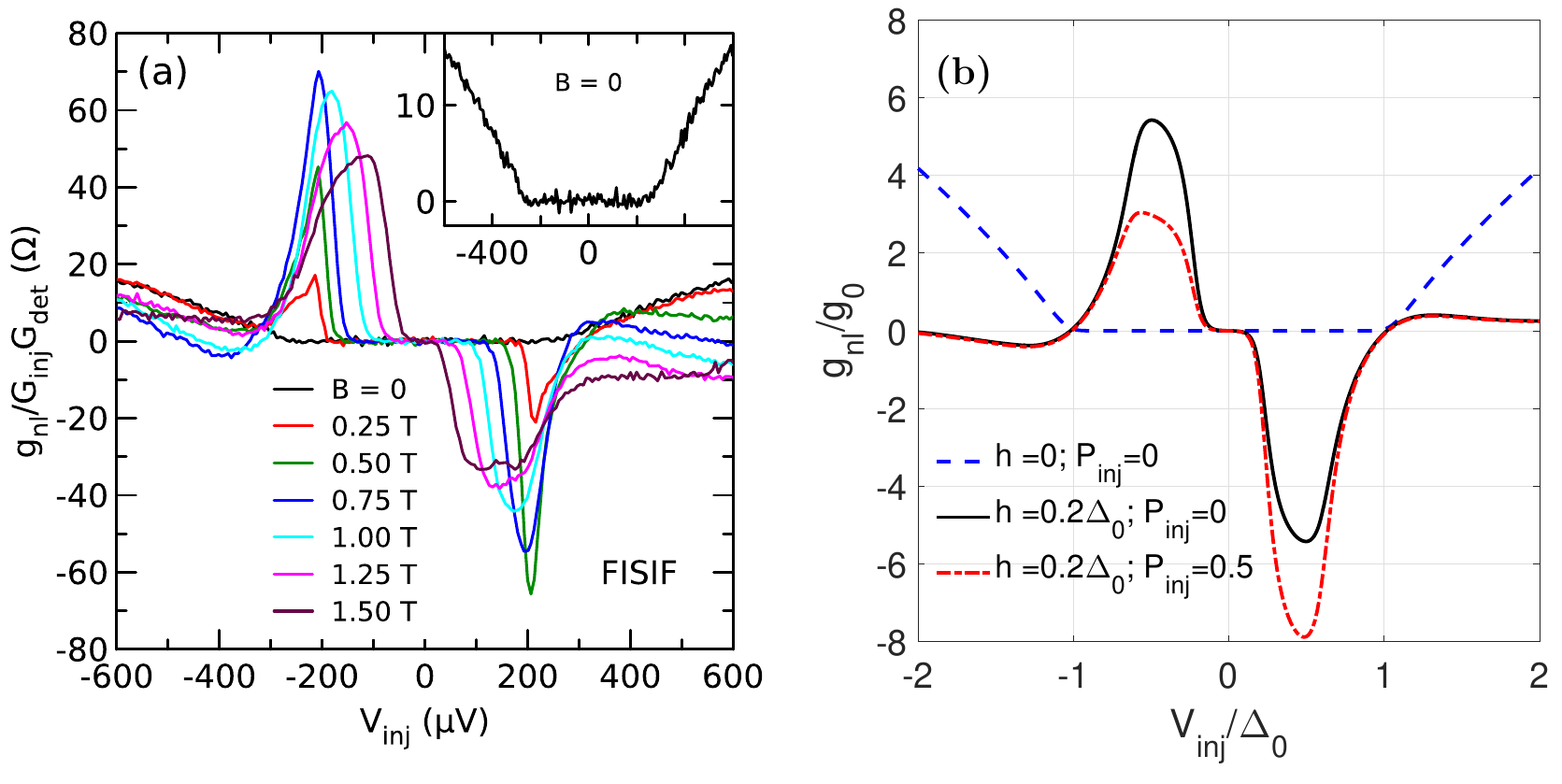}}
 \caption{\label{Fig:Cond-nl}
 (a) Nonlocal conductance measured as a function of the injecting voltage,  $g_{nl}(V_{\rm inj})$ adopted from \textcite{Hubler2012a}. 
 (b) The same quantity calculated using the kinetic theory for $\alpha_{orb}=1.33$, $\beta=0.5$, $(\tau_{sn} T_{c0})^{-1} =0.2$, $T=0.05 T_{c0}$,
 effective inelastic relaxation length $L=20\lambda_{sn}$, $L_{\rm det}=5\lambda_{sn}$. 
 Black solid and red dash-dotted curves correspond to the injection 
 from  non-ferromagnetic ($P_{inj}=0$) and ferromagnetic ($P_{inj}=0.5$) electrodes, respectively at the spin-splitting $h=0.2\Delta_0$. Blue dashed line corresponds to $h=0$. 
  The conductance is normalized to 
  $g_0 =  R_\xi/(R_{inj}R_{det})$,
  where $R_\xi = \xi/(A_s \sigma_N)$ is the normal-state resistance of the wire with length $\xi$ and cross section $A_s$. 
 }
 \end{figure}

Specifically, one of the setups studied   by \textcite{Hubler2012a}, was a lateral non-local spin valve {(see Fig.~\ref{fig:experiemnt_beckmann})} where the experimentalists determined the non-local differential conductance  \begin{equation}
  g_{nl}= \frac{d I_{\rm det}}{d V_{\rm inj}} \; .\label{eq:gnl0}
  \end{equation}
Typical experimental curves are shown in Fig.~\ref{Fig:Cond-nl}a, adapted from \textcite{Hubler2012a} and Fig.~\ref{Fig:Cond-nl}b
shows the results calculated from the kinetic equations.

If the detector is a ferromagnet with magnetization collinear with the spin accumulation in the wire, the current at the detector {for $V_{\rm det}=0$ is obtained from Eqs.~(\ref{eq:charge_current},\ref{eq:currents_at_interface}),}
\begin{equation}\label{Eq:ZeroCurrentYGen}
  I_{\rm det}= (\mu  +  P_{\rm det} \mu_z)/R_{\rm det},
\end{equation}
where 
$R_{\rm det}=R_\square/A$ is the detector interface resistance in the
normal state, $A$ is the cross-sectional area of the detector, $\mu$ is the charge imbalance and $\mu_z$ the spin imbalance defined in Eqs. (\ref{eq:def_mu},\ref{eq:def_mus}). According to the explicit expressions (\ref{Eq:ChPot0},\ref{Eq:ChPotZ}),  the full description of  the non-local current requires all four non-equilibrium modes.  
 
 Particularly interesting is the contribution from the  second term in the  r.h.s.  of Eq.~(\ref{Eq:ChPotZ}). It  is nonzero when the spin splitting described by $N_-$ is nonzero and it provides  a  qualitative explanation of 
 the experiments by \textcite{Quay2013,Hubler2012a,PhysRevB.87.024517}:
 The  spin imbalance $\mu_z$,  
being related to  the energy nonequilibrium mode 
 $f_L$, once excited can only  relax  via inelastic processes, especially mediated by the electron-phonon interaction.  
At low temperatures the corresponding decay length can be much larger than the spin decay length in normal metals.   
This explains  the long-range non-local signal observed in the experiments. 
The observed long-range spin
accumulation can thus be understood to result from the
 spin accumulation  generated by the  effective heating 
 of the superconducting wire caused by the  injection of nonequilibrium quasiparticles 
with energies larger than the superconducting gap \cite{silaev2015spin,silaev2015long,virtanen2016stimulated,bobkova2016injection,bobkova2015,krishtop2015-nst}. 
Such a heating can originate, for example, by an injected
current even from the non-ferromagnetic electrode. 
The heating is not sensitive to the sign of the bias voltage at the injector and hence  
  the generated spin imbalance must be an even function of the voltage, $\mu_z(V_{\rm inj})=\mu_z(-V_{\rm inj})$.  
 This leads  to an antisymmetric shape of the non-local spin signal in $g_{nl}$ with respect to $V_{\rm inj}$, {in agreement with the experimental observation \cite{wolf2014charge}.} All these features  occur only if the superconductor has a spin-split density of states induced  either by an external magnetic field or by the proximity to a ferromagnetic insulator.

{ A quantitative description of these effects can be provided by solving
 the kinetic equations for superconductors with 
 spin-split subbands \cite{silaev2015long}. In this case the diffusion couples non-equilibrium modes pairwise. In particular, the kinetic equations \eqref{eq:kineticeq} take the form}
 \begin{equation}\label{eq:KineticEquation}
\nabla\cdot\begin{pmatrix}
j_e\\ j_s\\ j_c\\ j_{se}
\end{pmatrix}=
\begin{pmatrix}
0& 0& 0& 0\\
0& 0& 0& S_{T3}\\
0& 0& R_T& R_{L3}\\
0& 0& R_{L3} & R_T+S_{L3}
\end{pmatrix}
\begin{pmatrix}
f_L\\ f_{T3}\\ f_T\\ f_{L3}
\end{pmatrix},
\end{equation}
where the spectral energy $j_e$, spin $j_s$,  charge $j_c$ and spin energy $j_{se}$ currents derived from the general Eq.(\ref{eq:general_current}) are
\begin{equation}\label{eq:Current1}
\begin{pmatrix}
j_e\\ j_s\\ j_c\\ j_{se}
\end{pmatrix}=
\begin{pmatrix}
D_L & D_{T3} & 0 & 0\\
D_{T3} & D_L & 0 & 0\\
0 & 0 & D_T & D_{L3}\\
0 & 0 & D_{L3} & D_T
\end{pmatrix}
\begin{pmatrix}
\nabla f_L\\\nabla f_{T3}\\\nabla f_T\\\nabla f_{L3}
\end{pmatrix}.
\end{equation}
Here $D_{L/T/T3/L3}$ are kinetic coefficients related to the spectral GFs \cite{silaev2015long}, $S_{T3/L3}$ are parts of  collision integrals describing spin relaxation, and $R_{T/L3}$ the coupling between the quasiparticles and the superconducting condensate.

On the one hand, the charge is coupled to the spin-energy mode [lower right block of Eq.~\eqref{eq:Current1}]. The relaxation of both of these modes, right hand side of Eq.~\eqref{eq:KineticEquation}, is nonvanishing for all energies, below and above the gap  due to the magnetic pair breaking effects \cite{schmid1975,nielsen1982pair}.  
On the other hand, the spin-splitting field couples the spin and energy modes,  $f_{L}$ and $f_{T3}$ respectively [upper left block of Eq.~\eqref{eq:Current1}]. As explained above, the energy mode 
can only decay  via inelastic scattering which at low temperature can be disregarded compared to the spin relaxation. 

Solutions of Eqs.~(\ref{eq:KineticEquation},\ref{eq:Current1}) along with Eqs.~(\ref{Eq:ChPot0},\ref{Eq:ZeroCurrentYGen}) reproduce the main features of the measured non-local conductance presented in Fig.~\ref{Fig:Cond-nl}a. Depending on the magnitudes of the spin-splitting field $h$ and the injector polarization $P_{\rm inj}$, we can identify three distinct parameter regimes affecting the symmetry of $g_{\rm nl}$. (i) When $h=P_{\rm inj}=0$ (blue dashed curve in Fig.~\ref{Fig:Cond-nl}b), the only contribution to the detector current comes from charge imbalance and $g_{\rm nl}$ is a symmetric function of the injection voltage. {In the absence of spin splitting and depairing effects, $R_T=0$ for $\varepsilon > \Delta$,} and hence charge imbalance decays only via inelastic scattering neglected here. {This explains the monotonic increase of $g_{nl}$ in Fig.~\ref{Fig:Cond-nl}b at large voltages}. (ii) For $P_{\rm inj}=0$ but in the presence of an applied field leading to $h\neq 0$ (black solid curve), charge relaxation is strongly enhanced due to the orbital depairing. The main long-range contribution comes from $\mu_z$ produced by the heating effect described above. The resulting $g_{\rm nl}$ is an antisymmetric function of $V_{\rm inj}$. (iii) When both $h\neq 0$ and $P_{\rm inj}\neq 0$ (red dash-dotted curve), an additional {symmetric long-range} contribution in $g_{\rm nl}$ results due to a thermoelectric effect at the injector. Note that in the case $h=0$, $P_{\rm inj} \neq 0$, there would be another symmetric contribution to $g_{\rm nl}$ due to the regular spin injection also present in the normal state. However, this is a short-range mode (decays via spin relaxation), and therefore does not show up beyond the spin relaxation length.

  { In the experiments by \textcite{Quay2013,Hubler2012a,wolf2014charge}   the spin-splitting field was caused  by an external magnetic field. Therefore   one needs to take into account the orbital depairing effect  of the magnetic field in addition to the Zeeman effect. The relative strength of the orbital depairing and the spin-splitting field is described by the dimensionless parameter $\alpha_{\rm orb}= (h\tau_{orb})^{-1}$. In Fig.~\ref{Fig:Cond-nl} we choose the value $\alpha_{\rm orb}=1.33$,  which should correspond  to the experiment by  \textcite{Hubler2012a}.}

In the presence of  a supercurrent, all coefficients of the matrix in Eq.~\eqref{eq:Current1} are nonzero \cite{aikebaier2017supercurrent}. As a result, for example the spin and charge modes are directly coupled by diffusion.

 \subsection{Spin Hanle effect}
 \label{sec:spinhanle}

 In  the previous sections we assume that all  magnetizations and  the applied field are collinear.  If one  lifts this assumption,
 the applied field leads to a precession of the injected spin 
 around the field direction. This is  the spin Hanle effect that {in the normal state} has been extensively  studied  in the literature and observed in several experiments \cite{PhysRevLett.55.1790,jedema:345,jedema2003spin,Villamor2015,yang2008giant}. The Hanle precession can  be measured via the non-local conductance in a setup such as the one shown in Fig. \ref{fig:experiemnt_beckmann}. The non-local measured signal oscillates and decays as a function of the amplitude of the applied field. 
 
Formally the Hanle effect is described by the first term on the r.h.s of Eq.~\eqref{eq:kineticeq}.
{Indeed, one can derive the Bloch-Torrey  transport equation \cite{Torrey1956}  for the magnetic moment  $\bm{m}(\varepsilon,x)={\rm Tr} (\tau_{3}\bm{\sigma}g^{K})/8$ from Eq.~\eqref{eq:kineticeq} \cite{silaev2015spin}. It reads
\begin{equation}
   \label{LL-1}
   \frac{\partial \bm m}{\partial t} + \nabla\cdot \bm j_{s} = 
    \gamma{\bm m}\times 
    {\bm h}_s - {\bm m}/\tau_S .
\end{equation}
Here $\gamma=-2$ is the electron gyromagnetic ratio and $\bm j_{s}$ is
the spin current density tensor.
In the normal state the spin relaxation  $\tau_S$ and Zeeman field ${\bm h}_s$ {are}  energy independent. This explains why  the {nonlocal resistance vs. field curve} does not depend either  on temperature or on the type of spin relaxation (magnetic  or spin-orbit impurities). In contrast, {they are predicted to be strongly energy dependent in the superconducting state, and the precession and decay of the nonlocal signal disappear at $T \rightarrow 0$, whereas the shape of the curves at  intermediate temperatures  depends on the type  of  spin relaxation \cite{silaev2015spin}.}
Experimental evidence of the Hanle effect in the superconducting state has not been reported {so far}. }

\subsection{Spin imbalance by ac excitation}
\label{sec:acdynamics}

The quasiparticle $f_{T,j}$ mode --- {or equivalently, the quasiparticle magnetic
moment $\bm{m}(\varepsilon,x)$ above} --- can be excited by an external
ac magnetic field, which via the Zeeman coupling generates a
conduction electron spin resonance
\cite{aoi1970-tes,maki1973-tes,vier1983-oce,yafet1984-ces,nemes2000-ces}.
This was recently studied experimentally in spin-split thin Al films
by \textcite{quay2015-qsr}.  As the $f_{T,j}$ mode can relax rapidly
via elastic spin-flip scattering, the linewidth seen in such
experiments is generally {$\tau_S^{-1}\simeq\tau_{sn}^{-1}$} instead of the time
scale of the long-ranged non-local spin signal.  Spin-flip scattering also
provides a channel via which electromagnetic fields can generate spin
imbalance through the orbital coupling
\cite{bentum1986-fia,virtanen2016stimulated}. For high enough driving
amplitude, the imbalance modifies the self-consistent $\Delta(T)$
relation, which develops additional features in the spin-split case
\cite{eliashberg1970-fss,virtanen2016stimulated}.  Effects related to
spin-splitting and relaxation can moreover be probed with tunnel
junctions at low frequencies \cite{quay2016-fdm} or via photoassisted
tunneling \cite{marchegiani2016-soq}.

\section{Thermoelectric effects\label{thermoel}}
Thermoelectric effects relate temperature differences to charge currents, and electrical potentials to heat currents. 
Thermoelectric effects are typically described via the linear response
relation between charge and heat currents $I$, $\dot Q$ and bias voltage and
temperature difference $V$ and $\Delta T$ across a contact:
\footnote{{In the case of thermoelectric effects, it is
    customary to talk about heat currents $\dot Q$ instead of energy
    currents $\dot U$, and we adapt this convention here. These are
    related by \cite{ashcroftmermin} $\dot
    Q=\dot U-\mu I/e$, where $\mu$ is a reference energy compared to the
    Fermi level. At linear response we can set $\mu=0$ in which case
    $\dot Q=\dot U$. On the other hand, when considering heat balance
    at non-vanishing voltages as in Sec.~\ref{sec_thermoelectric_sp},
    the two are not the same and rather the heat current 
    $\dot Q$ should be used.}}
\begin{equation}
\label{eq:thermoel}
\left(\begin{array}{c} I \\ \dot Q \end{array}\right) =
\left(\begin{array}{cc} G & \talpha
  \\ \talpha & G_{\rm th} T \end{array}\right) \left(\begin{array}{c} V \\ -\Delta
  T/T \end{array}\right).
\end{equation}
Here $G$ is the conductance and $G_{\rm th}$ the heat conductance of the contact. $\talpha$ is the thermoelectric coefficient.

 With a
 non-zero $\talpha$, electrical energy may be converted to heat or
cooling, or reciprocally a temperature difference may be converted to
electrical power. The efficiency of this conversion is typically described by the

thermoelectric figure of merit,

 \begin{equation}
 ZT  = \frac{\talpha^2}{G_{\rm th} G T -
   \talpha^2}=\frac{S^2 GT}{\tilde G_{\rm th}},
 \end{equation}
 where $S=\talpha/(GT)$ is the thermopower (Seebeck coefficient) and $\tilde G_{\rm
   th}=G_{\rm th}-\talpha^2/(GT)$ is the thermal conductance at a vanishing current. In particular, the maximum efficiency of a thermoelectric heat engine is \cite{snyder2003} ${\rm max} \eta = \eta_{\rm Carnot}
\frac{\sqrt{1+ZT}-1}{\sqrt{1+ZT}+1}$ with $\eta_{\rm Carnot}=\Delta T/T$. 
 Maximum efficiencies of the device are obtained when  $ZT\rightarrow \infty$.   At or above room temperature, the record-high figures of merit are
 obtained in certain strongly doped semiconductor structures  \cite{zhao2016,kim2015}. A typical record value for those cases is $ZT\gtrsim
 1\dots 2$.

The traditional view of thermoelectric effects in superconductors is that if they exist, they must be very weak. In bulk superconductors, this is partially because any thermoelectrically generated quasiparticle current is easily screened by a supercurrent \cite{meissner27thermoel}. 

Alternatively, one could then measure this supercurrent via an additional constraint to the phase of the superconducting order parameter in bimetallic multiply connected structures \cite{Ginsburg:1944vv}. However, even this thermally created phase gradient tends to be weak, owing to the near-complete electron-hole symmetry in superconductors.  \textcite{galperin74thermoel} showed that 
\begin{equation}
\alpha = \alpha_N G(\Delta/T),\quad G(x)=\frac{3}{2\pi^2} \int_x^\infty
\frac{y^2 dy}{\cosh^2(y/2)},
\label{eq:superalpha}
\end{equation}
where the latter form comes from the reduction of the quasiparticle
density in the superconducting state, and $\alpha_N$ is the
value of the thermoelectric coefficient in the normal state.
 The
 precise value of $\talpha_N$ depends on the exact electronic spectrum. For example, for a simple quadratic
dispersion $\talpha_N = \frac{\pi^2 G_T k_B^2 T}{6e E_F}$, where $E_F$ is the Fermi energy. At
 temperatures $T \ll \Delta/k_B$, $\talpha$ is thus
 expected to be a product of two small coefficients, $\talpha_N \propto
 k_B T/E_F$, and $G(\Delta/T)$. This is very small and not easy to measure quantitatively.    
 
However, superconductors do contain some ingredients for strong
thermoelectric effects, because the latter typically require strongly
energy dependent density of states of the charge carriers. This is
provided by the BCS density of states. Hence, if one can break the
electron-hole symmetry of the transport process via some mechanism,
superconductors can become very strong thermoelectrics. This is
precisely what happens in spin-split superconductors, as an exchange
field breaks the symmetry in each spin sector, but so that the overall
spin-summed energy spectrum remains electron-hole symmetric. Transport through a spin filter to a spin-split superconductor then can provide large thermoelectric effects because the two spins are weighed differently \cite{Ozaeta2014a,Machon2013,Machon2014}. 
We discuss these effects  in  this  section.

\subsection{Charge and heat currents  at a spin-polarized interface to a
  spin-split superconductor}\label{sec_thermoelectric_sp}

Consider a tunnel contact {from a non-superconducting reservoir R to a superconductor S in a spin-splitting
field.}  Let us assume that the
tunnel contact is magnetic, so that the conductance through it is
spin-polarized. Denoting the spin-dependent conductances in the
normal state by
$G_\uparrow,G_\downarrow$ we can parameterize them by the total
conductance $G_T=G_\uparrow+G_\downarrow$ and 
the spin polarization $
P=(G_\uparrow-G_\downarrow)/G_T$.
The total tunneling quasiparticle charge and heat currents are now expressed as a
sum over spin-dependent contributions, but otherwise of the standard form \cite{giaver1961,giazotto2006}. Denoting the spin-dependent
reduced density of states via $N_+=N_\uparrow+N_\downarrow$ and
$N_-=N_\uparrow-N_\downarrow$ the spin-averaged tunnel
currents { can be obtained from the Keldysh component of Eq.~\eqref{eq:currents_at_interface} after taking the corresponding traces:} 
\begin{align}
I&=\frac{G_T}{2e}\int_{-\infty}^\infty d\varepsilon  \left(N_++P
  N_-\right) (f_R-f_S)\label{eq:Ith}
  \\
\dot Q_i&=\frac{G_T}{2e}\int_{-\infty}^\infty d\varepsilon (\varepsilon-\mu_i) (N_+ + P N_-) (f_R-f_S).\label{th_qi}
\end{align}
Here $f_{R/S}=n_F(E-\mu_{R/S};T_{R/S})$, $n_F(E;T)=\{\exp[E/(k_B
 T)]+1\}^{-1}$ are the (Fermi) functions of the reservoirs biased at
 potentials $\mu_{R/S}$ and temperatures $T_{R/S}$.  The
reduced density of states in the superconductor for spin $\sigma$ is $N_\sigma(\varepsilon)$.  The heat current
$\dot Q_\sigma^i$ is calculated separately for $i=R,S$, using the
potential $\mu_{R/S}$, because the two heat currents differ by the
Joule power $I(\mu_R-\mu_S)/e$.  In  the analysis below, we disregard the spin relaxation effects on the density of states, because this
assumption allows for some analytically treatable limits and because it is a fair approximation in the case of often used Al samples.

{The heat current from R is a non-monotonous
function of voltage} even in the absence of spin polarization or
temperature difference. In
particular, for voltage $V=(\mu_R-\mu_S) \approx \Delta/e$, it is
positive, i.e., reservoir $R$ cools down \cite{nahum1996,leivo1996,pekola2004}. 
This
heat current is quadratic in the voltage, and therefore it does not
result from the usual Peltier effect [Eq.~\eqref{eq:thermoel} for
$\dot Q$] where the cooling power is linear
in voltage.

\begin{figure}
 \includegraphics[width=\columnwidth]{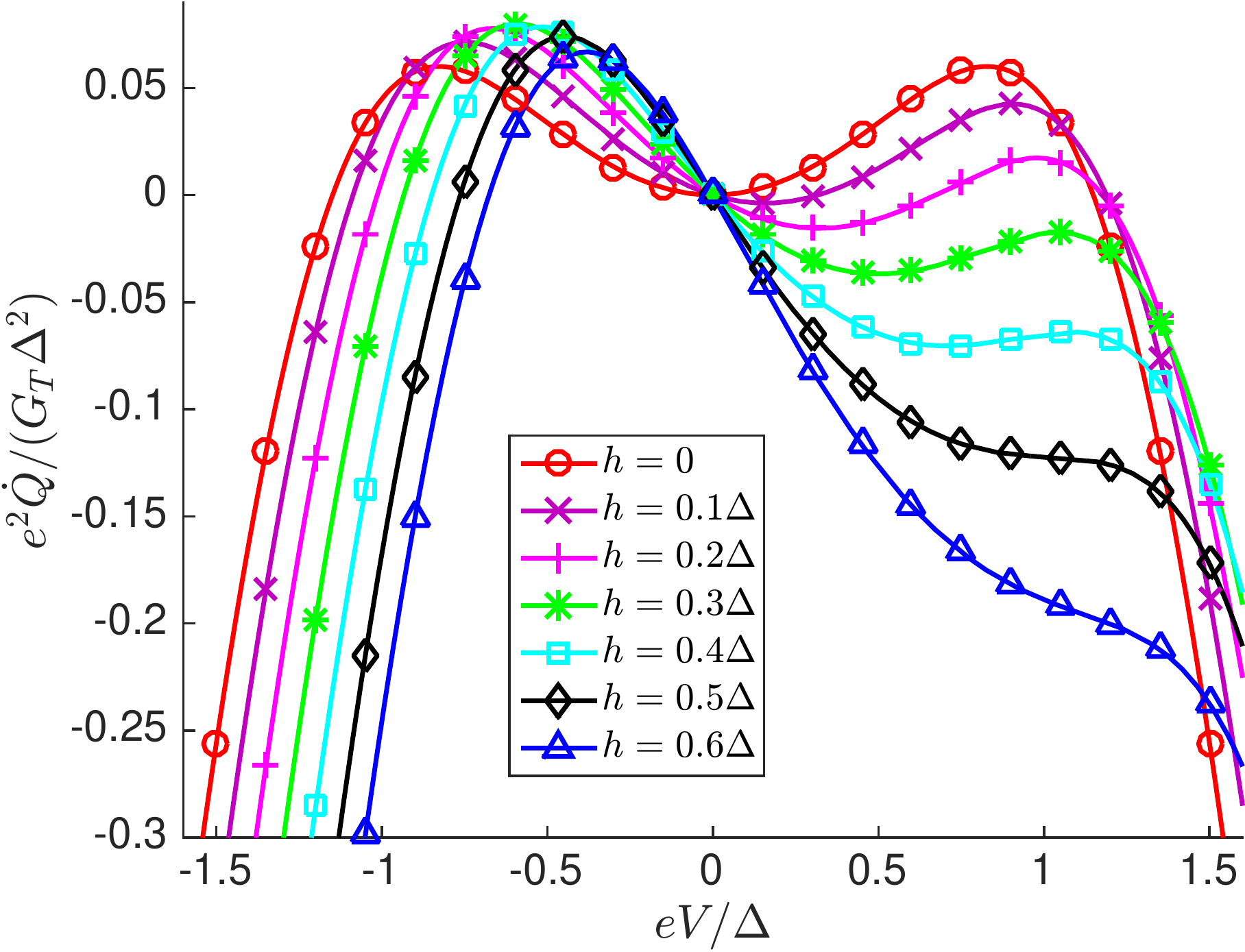}
 \caption{\label{fig:coolingpower}
  Cooling power from reservoir R vs.~voltage for different values of the
   exchange field $h$, assuming a unit polarization $P=1$ at the
   temperature $k_B T=0.3\Delta$ close to that yielding optimal cooling
   for $P=h=0$. The exchange
   fields are given in units of $\Delta$. Changing the sign of $P$ or $h$ 
   inverts the voltage dependence with respect to $V=0$.}
 \end{figure}

Interestingly, in the presence of spin
polarization $P$ and with a non-zero spin-splitting field $h$ in the superconductor, the cooling power is nonzero even in the linear response regime, {\it i.e.} low voltages \cite{Ozaeta2014a}. 
As an example we show in Fig.~\ref{fig:coolingpower} the cooling power from reservoir R as a function of voltage 
for various values of $h$, assuming the
ideal case of unit spin polarization $P=1$.

Contrary to the spin-independent case, the N-FI-S  element can also be used to refrigerate the superconductor.  Electron refrigeration using 
magnetic elements have been studied by \textcite{Rouco2017}.

\subsection{Linear response: heat engine based on a
  superconductor/ferromagnet structure}
\label{subs:linearresponseheatengine}

As can be seen in Fig.~\ref{fig:coolingpower}, the simultaneous presence of the non-vanishing spin
polarization $P$ and a spin-splitting field $h$ lead to a heat current
that has a linear component in the voltage $V$. This component is
nothing else than the Peltier effect.
In the limit $k_B T \ll \Delta-h$ the
linear-response coefficients evaluate to \cite{Ozaeta2014a}
\begin{align}
G&\approx G_T\sqrt{2\pi\tilde\Delta} \cosh(\tilde h)e^{-\tilde\Delta}
\,,\label{eq:NFISconductance}
\\
 G_{\rm th}&\approx \frac{k_B G_T\Delta}{e^2}\sqrt{\frac{\pi}{2\tilde\Delta}}e^{-\tilde\Delta}\left[e^{\tilde h}(\tilde\Delta-\tilde h)^2+e^{-\tilde h}(\tilde\Delta+\tilde h)^2 \right]
\,,
\\
 \talpha &\approx \frac{G_T P }{e}\sqrt{2\pi\tilde\Delta}e^{-\tilde\Delta}\left[\Delta\sinh(\tilde h)-h\cosh(\tilde h)\right]
\,,\label{eq:alpha}
 \end{align}
where $\tilde \Delta=\Delta/(k_B T)$ and $\tilde h=h/(k_B T)$. These
yield the thermopower 
\begin{equation}
S=\frac{\talpha}{GT} \approx \frac{P \Delta}{e
T}[\tanh(\tilde{h})-h/\Delta].
\end{equation}
At low temperatures the thermopower is maximized for $h=k_B T
{\rm arcosh}[\Delta/(k_B T)]$, where it is
\begin{equation}
S_{\rm max} \approx \frac{k_B}{e} P \left[\frac{\Delta}{k_B
    T}-{\rm arcosh}\left(\sqrt{\frac{\Delta}{k_B T}}\right)\right].
    \label{eq:thermopower}
\end{equation}
It can hence become much larger than the ``natural scale'' $k_B/e$,
and even diverge towards low temperatures. However, such a divergence
comes together with the vanishing of the conductance,
Eq.~\eqref{eq:NFISconductance}, and therefore is in practice either cut off by
circuit effects, where the impedance to the voltmeter becomes lower
than the contact impedance, due to spin relaxation neglected above, or alternatively by additional
contributions beyond the BCS model. The latter ones are described in
more detail by \textcite{Ozaeta2014a}. Nevertheless, with proper circuit
design one should be able to measure a thermopower much exceeding
$k_B/e$ in this setup.

\begin{figure}
\centering
\includegraphics[width=\columnwidth]{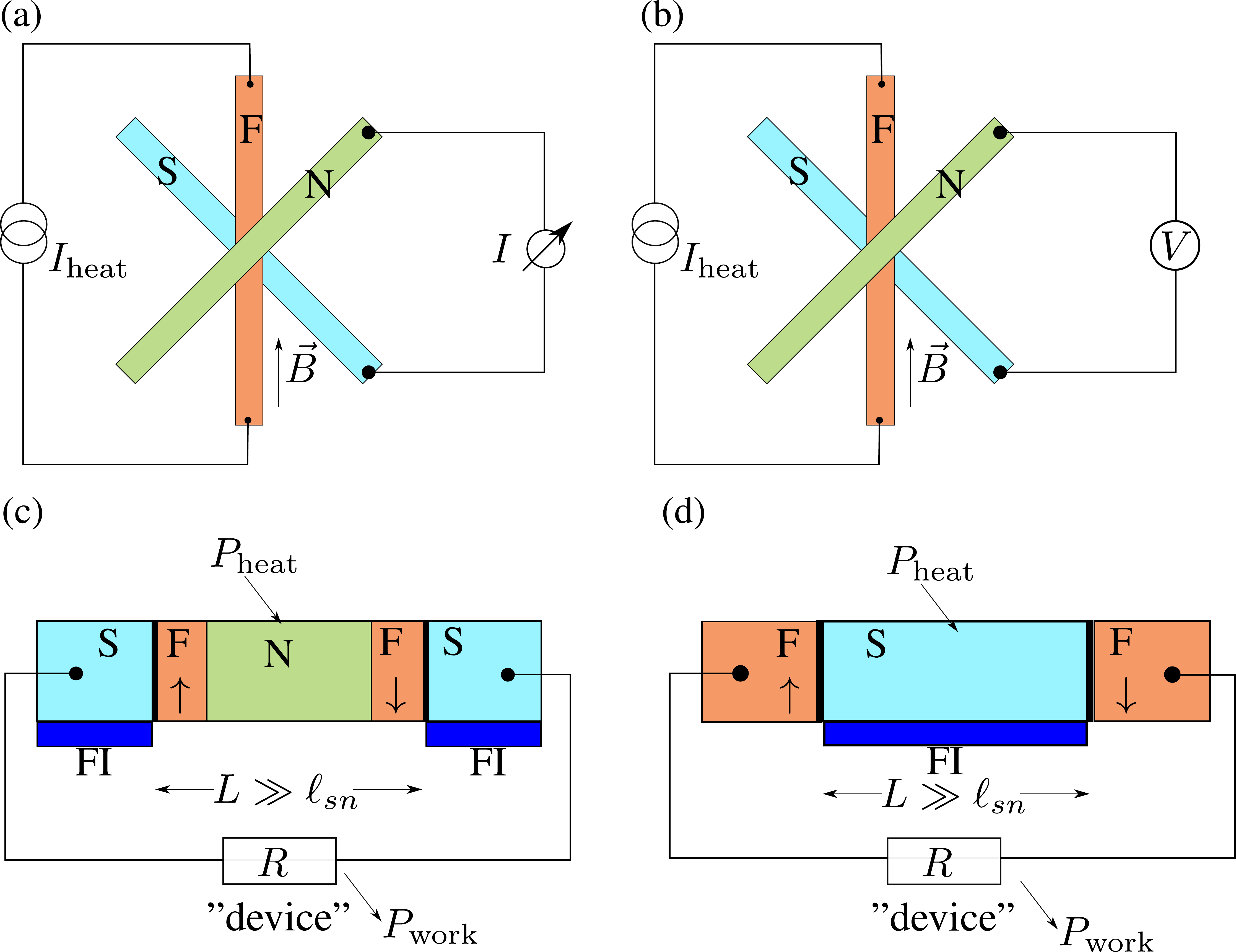}
\caption{\label{fig:beckmannexp} a) Schematic setup for measuring the thermoelectrically induced
  current, used by \textcite{Kolenda2016}. S, F, and N stand for a
  superconductor, ferromagnet and a normal metal, whereas FI is a
  ferromagnetic insulator. b) Setup used for a
  direct measurement of the Seebeck effect. c) Heat engine realized in
  a lateral setup with
  ``n-doped'' and ``p-doped'' junctions using a FNF trilayer with
  antiparallel magnetization directions. To disregard spin
  accumulation, the island has to be large compared to the spin
  relaxation length. d) Heat engine with a spin-split superconducting
  island. The ferromagnets can also be replaced by a
  normal metal if the interfaces to the superconductor contain a
  ferromagnetic insulator. In (c) and (d), the heating power $P_{\rm
    heat}$ is partially converted to ``useful'' work $P_{\rm work}$
  dissipated on the load.}
\end{figure}

\begin{figure}
\centering
\includegraphics[width=\columnwidth]{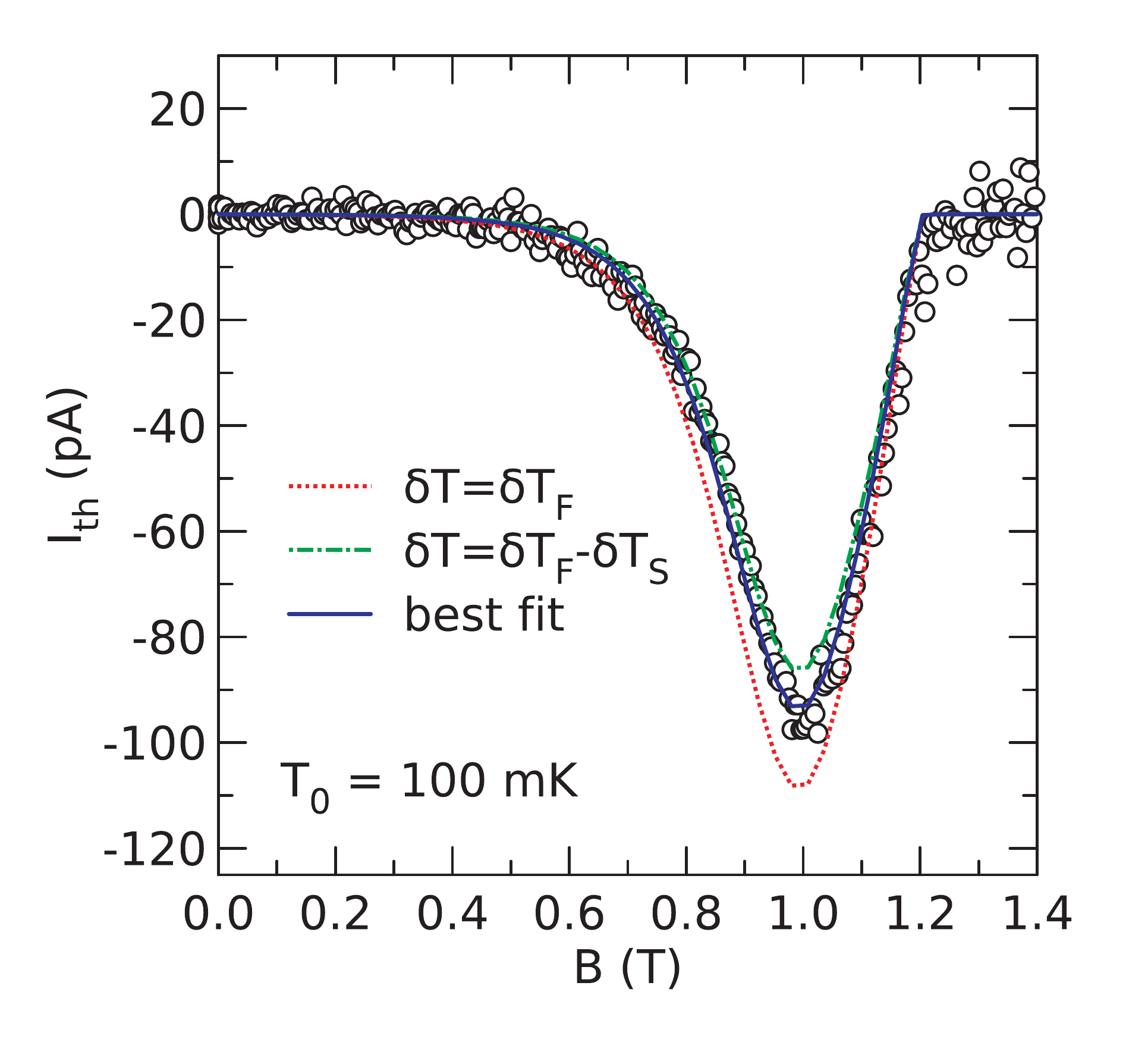}
\caption{\label{fig:beckmannthermoel} Thermoelectric current as a
  function of the applied magnetic field, measured in
  \cite{Kolenda2016}. The circles show the measurement values, the
  solid lines show a comparison to Eq.~\eqref{eq:Ith}. The three solid lines correspond to
  slightly different temperature differences; for further details, see
  \cite{Kolenda2016}. From \textcite{Kolenda2016}.
}
\end{figure}

The above theoretical predictions in the linear response regime were
confirmed experimentally by
\textcite{Kolenda2016,Kolenda2017}.
In particular, they prepared a sample containing a
crossing of three types of metals, a normal-metallic Cu, ferromagnetic
Fe, and superconducting Al. The measured configuration is  sketched in 
Fig.~\ref{fig:beckmannexp}a. The electrons in the ferromagnetic wire
were heated with the heater current $I_{\rm heat}$, producing a
temperature difference between the ferromagnet and the superconductor.
 The contact between the ferromagnet and the normal metal is ohmic and therefore the temperature difference between them is negligibly small.  
 Then the thermoelectric current was measured as a
function of the magnetic field $\vec B$ applied parallel to the
ferromagnetic wire. The agreement between the experimental results and
the above described tunneling theory was excellent (see Fig.~\ref{fig:beckmannthermoel}). The temperature difference between the ferromagnet and the superconductor was a fitting parameter, whereas the polarization  $P$ was fitted from the conductance spectrum.   In the experiment it was fitted to the value $P=0.08$, a modest value  attributed  to the  thin oxide barrier between the Fe and the Al layers.  In principle larger values of $P$ can be obtained  by increasing the thickness of the oxide  barrier  \cite{munzenberg2004superconductor}, but this of course  would  reduce the amplitude of the thermoelectric current.

In the experiment, the thermoelectric current was measured rather than
the voltage. In that case the impedance of the sample dominated that
of the measurement lines. This is why the
measurement yielded the exponentially low thermoelectric current,
which nevertheless was {sizeable}. The measurement
configuration in Fig.~\ref{fig:beckmannexp}b would have
directly measured the generated voltage
drop (\emph{i.e.}, Seebeck effect) instead of the current. This voltage
results from the ratio of two exponentially small functions, the
thermoelectric coefficient $\talpha$ and the conductance $G$, and
itself is not small. Such a measurement would then tell about spurious
effects, for example due to spin relaxation, or due to the presence of fluctuations or states
inside the gap. These effects would  limit the diverging Seebeck coefficient at low temperatures
\cite{Ozaeta2014a}.  Better still, replacing the normal metal with
another superconductor with an inverse spin-splitting field, would have
resulted to twice as large signal (corresponding to a series of p- and
n-doped thermoelectric elements), but would not be possible to create
as such with a magnetic field. The solution would be furthermore to
replace the ferromagnetic wire by an FNF heterostructure [Fig.~\ref{fig:beckmannexp}c, where the
ferromagnets have antiparallel magnetizations, for example due
to different coercive fields, and the normal metal N would serve as a spacer between
them]. To reach high figures of merit, the ferromagnetic metals should also be replaced by
ferromagnetic insulators, which can reach very high values of spin
polarization (see Table \ref{table:FIS}), with $P$ exceeding 0.9.

\begin{figure}
\includegraphics[width=\columnwidth]{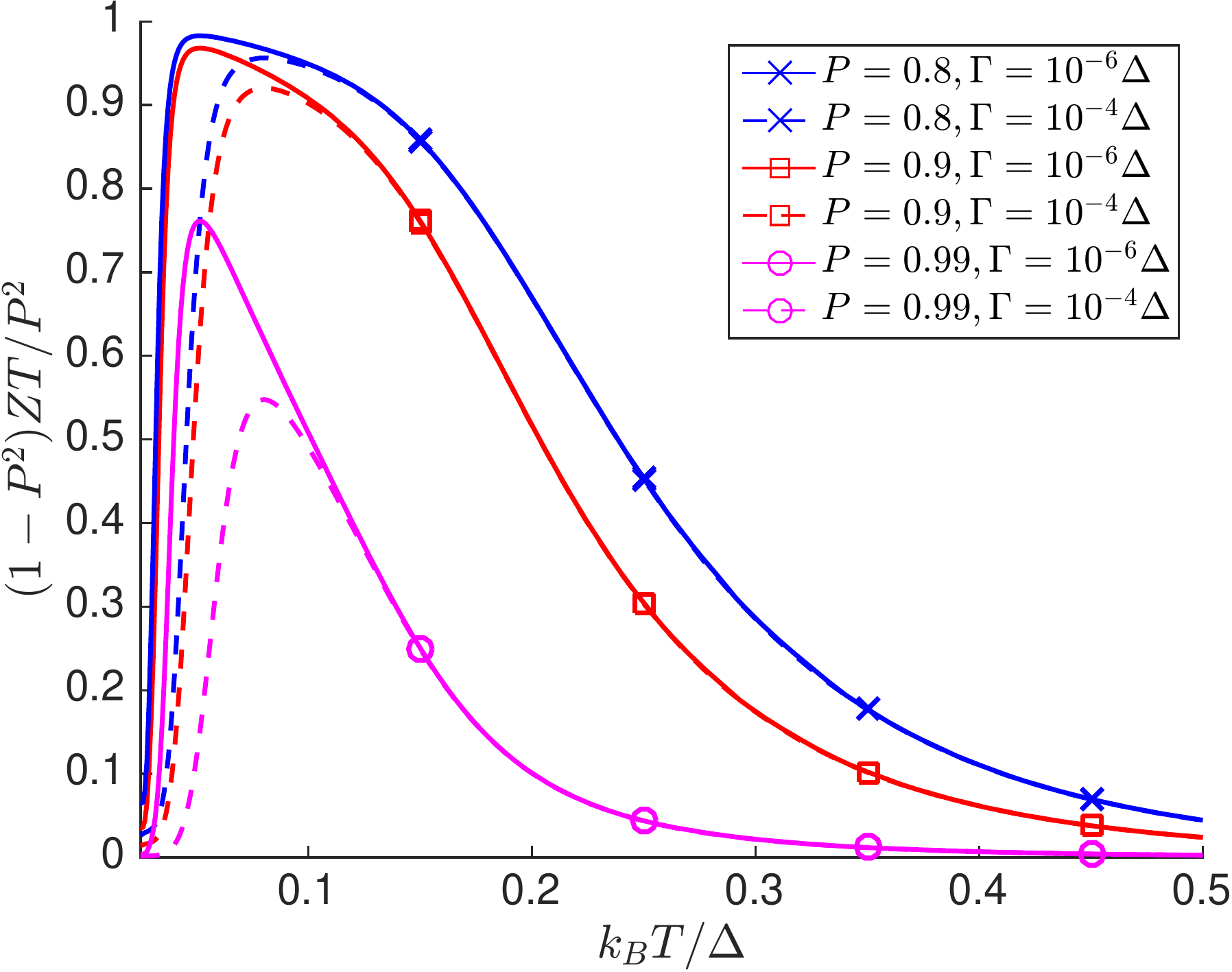}
\caption{\label{fig:ZTsuper2}
  Figure of merit in a N-FI-S-FI-N heat engine as a function of
  temperature for polarizations $P$ of the junction. The figure has been calculated with
  $h=0.5\Delta$ and $g=1000$, without calculating $\Delta$ self-consistently. The solid lines correspond to $\Gamma=10^{-6}\Delta$ and the dashed lines to $\Gamma=10^{-4}\Delta$. The
  figure of merit at low temperatures reaches very close to
  $P^2/(1-P^2)$ unless $P$ is very close to unity, but the
  exact temperature scale where this happens depends on the value of
  polarization. At the lowest temperatures $ZT$ is limited by another
  spurious heat conduction process, due to nonzero density of states
  inside the gap, described here by the Dynes $\Gamma$ parameter. }
\end{figure}

The island setup in Figs.~\ref{fig:beckmannexp}(c) and (d) also
realizes a thermally isolated structure, in contrast to those in (a)
and (b). This allows realizing a heat engine, where the voltage
measurement is replaced by the ``device'' to be powered with the
engine, with resistance that should be matched to the thermoelectric element. If only the electrons of the ferromagnetic island are heated, the main spurious heat conduction mechanism is due to electron-phonon coupling. In that case it is advantageous to use the structure (d), because the electron-phonon heat conductance is weaker in a superconductor \cite{kaplan1976,heikkilaup2017} than in a normal metal \cite{wellstood1994}. For example, Fig.~\ref{fig:ZTsuper2} shows a prediction for the resulting temperature dependence of the thermoelectric figure of merit ZT in structure (d), including this spurious heat conduction. In an optimized structure, very large ZT could thus be expected. In the picture, $g=5 k_B^5 \sqrt{2\pi} e^2 \Sigma \Omega\Delta^3/(2 G_T)$ is a dimensionless quantity characterizing the
relative strength of electron-phonon coupling (characterized by $\Sigma$ \cite{giazotto2006}) to the tunnel coupling of the thermoelectric element in an island with volume $\Omega$. For
example, for $\Omega=0.005$ $\mu$m$^3$, $\Sigma=10^{9}$ W
$\mu$m$^{-3}$K$^{-5}$ and $1/G_T=30$ k$\Omega$, $g=1000$.

Note that it is really the presence of the spurious electron-phonon heat conduction that limits the highest available values of $ZT$. Often such  spurious mechanisms are disregarded from the theoretical analysis, for example in the case of quantum dots \cite{hwang16}.

{Even if the true figure of merit of the type of heat engine discussed
  above can be made high, these systems cannot obviously be used to
  replace room-temperature thermoelectric devices to be applied for
  example in energy harvesting. However, there are other applications
  where the large figure of merit may turn out to be essential. For
  example, this type of thermoelectric heat engine can be used for
  thermal radiation sensing at low temperatures
  \cite{giazotto2006,heikkilaup2017}. Another possible use of the
  thermoelectric effects would be in non-invasive low-temperature thermometry \cite{giazotto2015b}, where
  the temperature (difference) profiles could be read from the
  thermopower, without having to apply currents. In a scanning mode
  this would hence be a low-temperature version of the method used by \textcite{menges16}.}

Note that the above discussion disregards the effect of spin-orbit or spin-flip scattering on the superconducting state. It  limits $ZT$ especially in heavy-metal superconductors. The associated effects were considered by \textcite{bergeret2017nonequilibrium,rezaei2017spin}.

\subsection{Spin Seebeck effect \label{subs:spinseebeck}}

Besides the large thermoelectric effect, the
 contact between spin-split superconductors with other conducting materials can exhibit a large (longitudinal) spin Seebeck effect, where a temperature difference drives spin currents to/from the spin-split superconductor \cite{Ozaeta2014a}. In this case the charge, heat, spin and spin heat currents are described by the full  {\cite{onsager31,jacquod12,Machon2013}} Onsager linear-response matrix
    \begin{equation} \label{Eq:Onsager}
   \left( \begin{array}{ccc}
   I \\ \dot Q \\ I_s \\ \dot Q_s
  \end{array} \right)  
    = \left(
  \begin{array}{cccc}
    G & \alpha & PG & \tilde \alpha  \\
    \alpha & G_{\rm th}T & \tilde \alpha & PG_{\rm th}T \\
    PG & \tilde \alpha & G & \alpha \\
    \tilde \alpha & PG_{\rm th}T & \alpha & G_{\rm th}T \\
  \end{array}
  \right)
  \left(\begin{array}{ccc}
    V \\ - \Delta T/T   \\ V_s/2 \\ - \Delta T_s/2T
  \end{array} \right), 
  \end{equation}   
where for $k_B T \ll \Delta-h$ the coefficients $G$, $G_{\rm th}$ and
$\alpha$ are given in Eqs.~(\ref{eq:NFISconductance}-\ref{eq:alpha}),
and $\tilde \alpha=\alpha/P$. Here $V_s$ and $\Delta T_s$ refer to
spin-dependent biases \cite{bergeret2017nonequilibrium}.
 
 The spin currents induced in the case of two spin-split superconductors, and the additional effects of Josephson coupling, magnetization texture and spin-orbit effects are discussed by \textcite{linder2016,bathen2017}. When either of the two materials realizes an island, the spin current can be converted into a spin accumulation $\mu_z$ that is determined from  the balance between thermally
induced spin currents and spin relaxation within the island. The above discussion on heat engines assumes a structure size much longer than the spin-relaxation length, and hence disregards this spin accumulation. The effect of the thermally induced spin accumulation on the superconducting gap was considered by \textcite{bobkova2017}, who predicted the associated changes in the critical temperature.

This spin Seebeck effect should be contrasted to the analogous phenomenon discussed in non-superconducting materials \cite{Uchida2014}. There, {a major contribution to the spin Seebeck signal is due to the thermally induced spin pumping  \cite{hoffman2013}}.

\subsection{Thermophase in a S(FI)S contact}
\label{sec:thermophase}
The large thermoelectric effect described {above} allows  for a large thermally induced phase gradient. 
This  was
theoretically investigated by \textcite{giazotto2015}. The total current in this case
consists of the sum of a thermoelectric current $I_{\rm th}$ and the supercurrent,
\begin{equation}
I=I_{\rm th} + I_c \sin(\varphi),
\end{equation}
where $I_{\rm th}$ is obtained from \eqref{eq:Ith} and $I_c$ is the
critical current for the junction with a phase
difference $\varphi$ of the order parameters across the contact.  The critical current is proportional to $\sqrt{1-P^2}$ \cite{Bergeret2012} and depends on the spin-splitting field in S \cite{bergeret2014}

In an electrically open configuration, the two currents must cancel,
and instead a {\it thermophase} $\varphi^{\rm th}$ develops across the
junction. This is obtained from
\begin{equation}
\sin(\varphi^{\rm th}) = -\frac{I_{\rm th}}{I_c}.
\label{eq:thermophase}
\end{equation}
The thermophase can be detected using a bimetallic loop with two
contacts, characterized by critical currents $I_{c1,2}$ and
thermophases $\varphi_{1,2}^{\rm th}$. For non-zero exchange field and
spin polarization $P$, the resulting thermophases can be much larger
than in ordinary bulk superconductors. Hence the temperature
dependence of the inductances play a more minor role {than in the case of superconductors without spin splitting \cite{vanharlingen80thermoel,Shelly16}}. For junctions
with non-equal thermophases and for negligible loop inductance (in
practice, $2e L I_{c1,2} /\hbar \ll 1$) in the absence of an external
flux the circulating current is
\begin{equation}
I_{\rm circ} = \frac{I_{c1} I_{c2}}{I_{c1}+I_{c2}}
\left[\sin(\varphi^{\rm th}_1)-\sin(\varphi^{\rm th}_2)\right].
\end{equation}
In the case of symmetric junctions both thermophases are the same and
the circulating current in the absence of an external flux
vanishes. However, as discussed by \textcite{giazotto2015}, the
thermoelectric current affects the response of the circulating current
to the external flux, allowing for their measurement also in that
case. 

Equation \eqref{eq:thermophase} requires that both sides of the
equation have an absolute value of at most unity, i.e., $|I_{\rm th}|
< I_c$. For a very large thermoelectric current, its cancellation with
a supercurrent is no longer possible, and instead a voltage across the
contact forms. In this case the direct current response of the junction is more similar to
the case discussed above in the linear response limit for a N-FI-S
junction. This regime was investigated in detail by
\textcite{linder2016}. Moreover, the nonvanishing dc voltage across the
superconducting junction leads to Josephson oscillations at the
frequency $2eV/h$, where $h$ is the Planck constant. Hence, the device
can be used as a temperature (difference) to frequency converter as discussed in
more detail by \textcite{giazotto2015b}.

\section{Summary and Outlook}\label{sec:outlook}

This review focuses on transport and thermal properties of superconducting hybrid structures with a spin-split density of states.   Such a splitting can be achieved either by an external magnetic field, or, more interestingly,  by placing  a ferromagnetic insulator (FI)   adjacent to a superconducting layer (S) (Sec.~\ref{sec-superwithh}).  
We discuss  several  experimental situations  with the help 
of a theoretical  framework (see Sec.~\ref{sec:Usadel} and \ref{sec:noneq_quasiclas}) based on the quasiclassical formalism, with which one can account for both thermodynamical and  nonequilibrium  properties of  such hybrid structures.  In order to account for effects beyond quasiclassics, as for example strong spin polarization, we combine the quasiclassical equations with effective boundary conditions.

Out-of equilibrium superconductivity by itself leads to a decoupling between the charge and energy degrees of freedom of the electronic transport.  
In this review we show that the combination between superconductivity and magnetism requires on one hand a description of additional nonequilibrium modes, spin and spin energy, and on the other
 hand to couples them all. This leads to novel and intriguing phenomena discussed in this  review with direct impact in latest research activities and proposed future technologies  based on superconductors and spin dependent-fields \cite{eschrig2011spin,eschrig2015spin,linder2015superconducting}.
By  using the theoretical formalism presented in this review one can predict and explain phenomena such as  the spin injection and relaxation (Sec.~\ref{spininjection}) in  superconductors  with an intrinsic exchange field along with their consequences  in  the transport properties.  We also discuss  a number of  striking thermoelectric effects in  superconductors  with a spin-splitting field (Sec.~\ref{thermoel}).

The best scenario for the phenomena and applications discussed here, and in particular for the thermoelectric effects,  
are FI-S systems where the spin splitting can be achieved  without the need of an applied magnetic field.
Hence it becomes important to look for ideal FI-S  material combinations. So far europium chalcogenides (EuO, EuS and EuSe) together with Aluminum films have shown large splittings and hence these are the best combination.  In addition, thin films of EuO or EuS can be used as almost perfect spin filters (see Table \ref{table:FIS}) and hence they  are good  candidates for realizing the near-optimal heat engines proposed  in Sec.~\ref{thermoel}.  One of the main challenges  from this perspective is to find FI-S combinations with large superconducting critical temperature and simultaneously a large spin splitting. Superconductors like Nb or Pb  on the 
one hand increase  $T_C$ with respect to  Al-based  structures, but on the other the spin-orbit coupling may spoil the sharp splitting as discussed in Sec.~\ref{sec:SCwithh}.  Recent experiments on GdN-NbN suggest large splittings \cite{PhysRevB.92.180510} but further research in this direction is needed.

In Sec.~\ref{sec:acdynamics} we  {briefly}  discuss  the dynamics of  spin-split superconductors in rf fields. Historically, magnetic resonance effects in superconductors are well studied, but fewer experiments have probed spin-split thin films.

Besides the effects discussed in this review, several theoretical studies  made striking predictions  in  mesoscopic systems with spin-split superconductors, such as the   creation of highly polarized spin currents \cite{absolute_spin_valve,PhysRevB.77.132501,giazotto2013quantum},  large supercurrents in FI-S-I-S-FI junctions \cite{bergeret2001enhancement},  junctions with switchable current-phase relations \cite{strambini2015mesoscopic}, and an almost ideal heat valve based on S-FI elements \cite{Giazotto:2013ei}. 

Although many of the transport phenomena in spin-split superconductors are now well-understood, we foresee a number of exciting avenues for future research. 

One further perspective of the present work is the extension of the Keldysh quasiclassical  formalism  used in  this review to include magneto-electric effects associated with the spin-orbit coupling (SOC). For a linear in momentum  SOC the generalization of this can be done by introducing an effective SU(2) gauge potential. The quasiclassical equations in this case  have been derived by \textcite{Bergeret:2013il,bergeret2014spin,bergeret2016manifestation}.  Effects such as the spin-Hall and  spin-galvanic effect in superconductors  have been studied in the equilibrium case \cite{konschelle2015theory}. Extending these results to a nonequilibrium situation, and also to time-dependent fields, would be an interesting further development and would allow for a detailed study of the well-controlled non-linearities associated to these effects in superconductors.  First steps in this  direction have been taken in \cite{espedal2017}.

Recent discoveries of 
skyrmionic states in chiral magnets \cite{
Nagaosa2012}
 have attracted a lot of attention due to the effects resulting from the interplay of magnetism and SOC \cite{Soumyanarayanan2016} which can induce chiral Dzyaloshinskii-Moriya interactions between magnetic moments.
Currently it is very interesting to study these effects in the presence of the 
 additional component --- superconductivity, when the
 exchange interaction is mediated by the Cooper pairs \cite{DEGENNES196610}. One can expect that in such systems superconductivity can induce  a non-trivial magnetic ordering and dynamics. These effects can show up in various systems including  ferromagnet/superconductor bilayers, surface magnetic adatoms and bulk magnetic impurities inducing the localized Yu-Shiba-Rusinov states modified by the SOC \cite{PhysRevLett.115.116602}.

Superconducting structures with strong spin-orbit coupling and exchange fields are also of high interest in view of engineering a platform for realization of topological phases and Majorana bound states \cite{qi2011-tia,beenakker2013-smf,alicea2012,hasan2010-cti}. Understanding and controlling the behavior and relaxation of nonequilibrium quasiparticles in these systems is also of importance, not least because of their influence on the prospects of solid-state topological quantum computation \cite{nayak2008}.

This review focuses exclusively on the nonequilibrium properties of superconductors in proximity to magnets. We expect the inclusion of the magnetization dynamics and its coupling to the electronic degrees of freedom via the reciprocal effects of spin transfer torque and spin pumping \cite{tserkovnyak2005} in the far-from equilibrium regime to lead to completely new type of physics, as the two types of order parameters affect each other. The coupling of supercurrent on magnetization dynamics and texture has been studied during the past decade \cite{waintal2002,houzet2008,richard2012-aci}, but the work where both systems are out of equilibrium has been mainly concentrated on Josephson junctions \cite{holmqvist2011,hikino2011,mai2011,holmqvist2014,kulagina2014} and much less attention has been paid to quasiparticle effects \cite{skadsem2011,linder2012,trif2013-dme}.

{Besides the rich physics offered by spin-split superconductors, they have been long used as tools to characterize equilibrium properties of magnets, especially their spin polarization. In this review (see end of Sec.~\ref{subs:linearresponseheatengine}) we outline two further possibilities related to their large thermoelectric response: accurate radiation sensing and non-invasive scanning thermometry. We believe there are also many other avenues to be uncovered, opened by the possibility for realizing a controlled combination of magnetism and superconductivity.}

\acknowledgments We thank Faluke Aikebaier, Marco Aprili, Detlef Beckmann,  Wolfgang Belzig, Irina Bobkova, Alexander Bobkov,  Matthias Eschrig, Yuri
Galperin, Francesco Giazotto, Vitaly
Golovach, Kalle Kansanen, Alexander Mel'nikov, Jagadeesh Moodera,  Risto Ojaj\"arvi,
Asier Ozaeta, Charis Quay, Jason Robinson,  Mikel Rouco, and Elia Strambini   for useful discussions. This work was
supported by the Academy of Finland Center of Excellence (Project No. 284594), Research Fellow  (Project No. 297439) and Key Funding (Project No. 305256) programs, the European Research Council (Grant No. 240362-Heattronics), the Spanish Ministerio de Econom\'ia y Competitividad (MINECO) (Projects No. FIS2014-55987-P and FIS2017-82804-P),
the European Research
Council under the European Union's Seventh Framework Program
(FP7/2007- 2013)/ERC Grant agreement No. 615187-COMANCHE.

\bibliography{superferro}

\appendix

\end{document}